\documentclass[fleqn,10pt]{wlscirep}
\usepackage[utf8]{inputenc}
\usepackage[T1]{fontenc}
\usepackage{color,soul}
\usepackage{mathptmx}
\usepackage{float}
\usepackage{amsmath}
\usepackage{amsmath}
\usepackage{blindtext}
\usepackage{gensymb}
\usepackage{booktabs,chemformula}
\usepackage{sidecap}
\usepackage{verbatim}
\usepackage{graphicx}
\usepackage{tablefootnote}
\usepackage{footnote}
\usepackage{xr}

\usepackage{dcolumn}
\usepackage{bm}
\usepackage{times}
\usepackage{subfigure}
\usepackage{lineno}
\usepackage{hyperref}
\usepackage{color}
\usepackage{babel}

\title{Electrically reconfigurable phase-change transmissive metasurface}

\author[1*]{Cosmin Constantin Popescu}
\author[2,*,$^{\ddag}$]{Kiumars Aryana}
\author[2]{Parth Garud}
\author[1]{Khoi Phuong Dao}
\author[3]{Steven Vitale}
\author[3]{Vladimir Liberman}
\author[4]{Hyung-Bin Bae}
\author[4]{Tae-Woo Lee}
\author[5]{Myungkoo Kang}
\author[5]{Kathleen A. Richardson}
\author[6]{Carlos A. R\'ios Ocampo}
\author[1]{Yifei Zhang}
\author[1,7]{Tian Gu}
\author[1,7]{Juejun Hu}
\author[2,$^{\dagger}$]{Hyun Jung Kim}

\affil[1]{Department of Materials Science and Engineering, Massachusetts Institute of Technology, Cambridge, 02139, MA, USA}
\affil[2]{NASA Langley Research Center, Hampton, 23666, VA, USA}
\affil[3]{Lincoln Laboratory, Massachusetts Institute of Technology, Lexington, MA, 02421, USA,}
\affil[4]{KAIST Analysis Center, Korea Advanced Institute of Science and Technology, Yuseong-gu, Daejeon 34141, Korea}
\affil[5]{CREOL, The College of Optics \& Photonics University of Central Florida Orlando, FL, 32816, USA}
\affil[6]{University of Maryland, Department of Materials Science \& Engineering, College Park, MD, USA}
\affil[7]{Materials Research Laboratory, Massachusetts Institute of Technology, Cambridge, 02139, MA, USA}

\affil[*]{These authors contributed equally to this work.}
\affil[$\ddag$]{kiumars.aryana@nasa.gov}
\affil[$\dagger$]{hyunjung.kim@nasa.gov}

\begin{abstract}
Programmable and reconfigurable optics hold significant potential for transforming a broad spectrum of applications, spanning space explorations to biomedical imaging, gas sensing, and optical cloaking. The ability to adjust the optical properties of components like filters, lenses, and beam steering devices could result in dramatic reductions in size, weight, and power consumption in future optoelectronic devices. Among the potential candidates for reconfigurable optics, chalcogenide-based phase change materials (PCMs) offer great promise due to their non-volatile and analogue switching characteristics. Although PCM have found widespread use in electronic data storage, these memory devices are deeply sub-micron-sized. To incorporate phase change materials into free-space optical components, it is essential to scale them up to beyond several hundreds of microns while maintaining reliable switching characteristics. This study demonstrated a non-mechanical, non-volatile transmissive filter based on low-loss PCMs with a 200 $\mu$m$ \times $200 $\mu$m switching area. The device/metafilter can be consistently switched between low- and high-transmission states using electrical pulses with a switching contrast ratio of 5.5 dB. The device was reversibly switched for 1250 cycles before accelerated degradation took place. The work represents an important step toward realizing free-space reconfigurable optics based on PCMs.

\end{abstract}

\begin{document}
\flushbottom
\maketitle

\renewcommand{\thefootnote}{\roman{footnote}}

\section{Introduction} 
Unlike traditional optics whose properties are fixed after fabrication, reconfigurable optics enable agile tuning of their functions to dynamically adapt to different needs. The applications for reconfigurable optics span adaptive optical microscopy to data communications\cite{horst2023tbit} and observation of celestial bodies in space telescopes\cite{hampson2021adaptive,davies2012adaptive}. Thus far, various techniques have been adopted for real-time manipulation of electromagnetic wave properties (amplitude, phase, and polarization)\cite{salter2019adaptive} either for focusing, filtering or steering light such as liquid crystals\cite{algorri2019recent}, deformable mirrors\cite{madec2012overview}, filter wheels\cite{beichman2012science}, and digital mirror devices\cite{ren2015tailoring}. While each of these methods provides advantages within their specific applications, there remains a pressing demand for a compact, non-mechanical, non-volatile beam-shaping technology capable of continuously tuning light properties. Such a technology should also be compatible with standard microfabrication processes to enable its rapid and cost-effective integration with other technologies\cite{gyger2021reconfigurable, gu2023reconfigurable}.

\begin{figure*}
\centering
{\includegraphics[scale= 0.53]{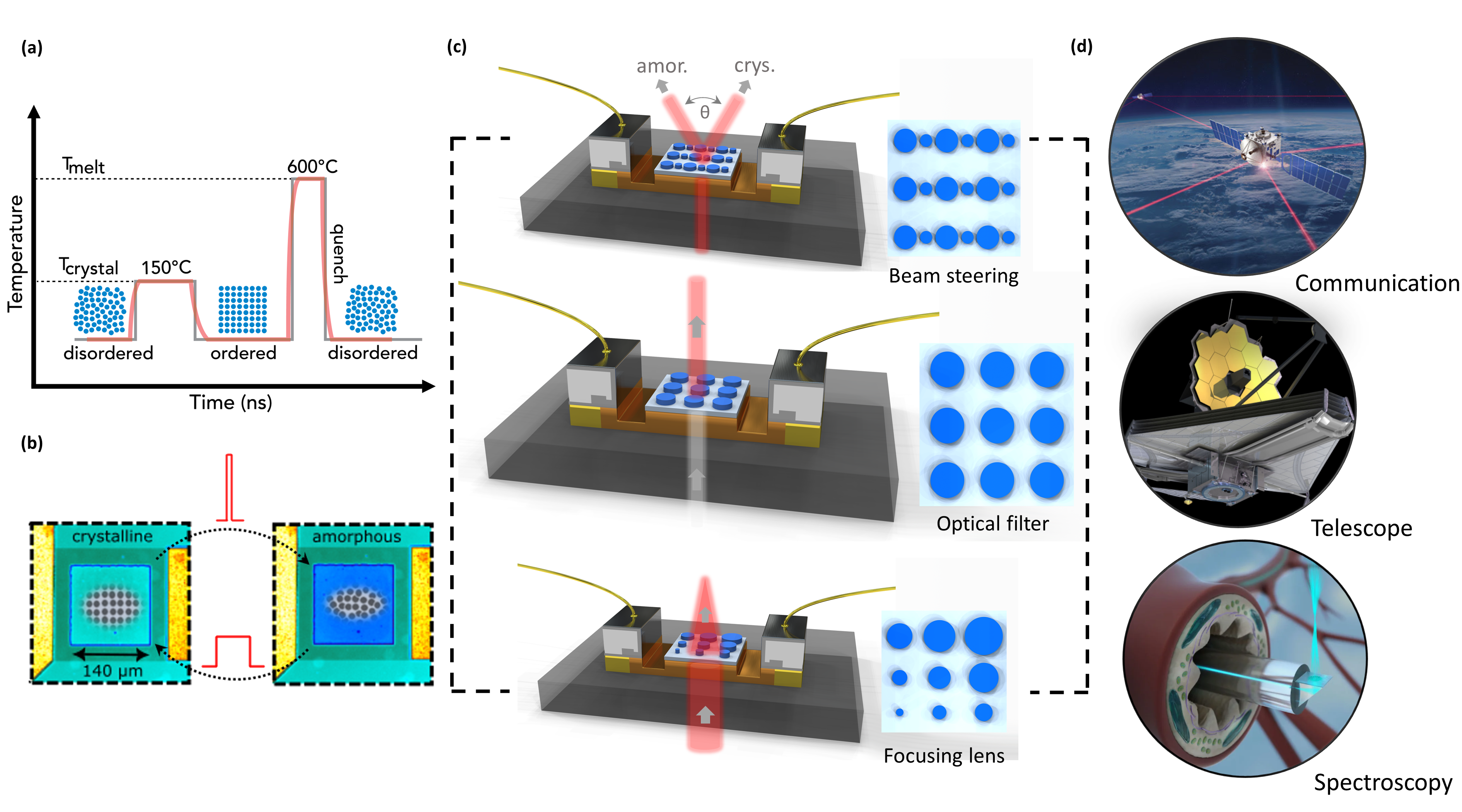}}
\caption{Concept of the operation of a PCM based device, showing temperature of the device vs time (a) along with NIR reflection images of a PCM film after amorphization on a doped SI heater (b), types of beam manipulations highlighting the types of meta-atoms required as a minimal basis for the individual operation (c) and potential areas of application for each individual type of beam manipulation (d). 
\label{fig_01}}
\end{figure*}

Active metasurfaces have emerged as a new avenue for creating compact, high-performance reconfigurable optical devices that can be electrically controlled without the need for mechanical components\cite{zhang2021electrically, abdollahramezani2022electrically, Wu2019DynamicMetasurfaces, Iyer2018UniformMetalenses, Kim2019PhaseMetasurfaces, Yin2017BeamMetasurfaces, julian2020reversible, raeis2017metasurfaces, de2020reconfigurable, wang2023varifocal}. This innovative method for controlling photon propagation within a highly compact configuration has led to the development of small form factor optical components that were previously only achievable through sophisticated, bulky optical systems\cite{shaltout2019spatiotemporal}. By incorporating an active material, such as a chalcogenide-based phase change alloy, into these metasurfaces architecture, a large degree of tunability can be achieved through reversible phase transformation. Initiating a phase transformation in PCMs necessitates the application of a strong thermal excitation to rapidly heat the PCM and raise its temperature above the crystallization or melting point as depicted in Fig. \ref{fig_01}(a). In the case of electrothermal switching\cite{wuttig2017phase, zhang2021myths}, the thermal excitation is applied through an on-chip resistive micro-heater, yielding a form factor suited for compact chip-scale integration. For instance, an almost 400\% reflectance change upon phase transformation was demonstrated in a PCM metasurface integrated on a metal micro-heater \cite{zhang2021electrically}. In another work, a remarkable 11-fold change in surface reflectance was achieved in a plasmonic metasurface similarly fabricated on metal micro-heaters\cite{abdollahramezani2022electrically}.

These devices however are limited to operating in the reflective mode due to the use of opaque metal heaters. Moreover, endurance (lifetime) of the devices have been limited to approximately 50 cycles in the reports. Optically transparent heaters are essential to enable next-generation active PCM metasurface devices suitable for transmissive optics. Doped crystalline Si features low loss in the infrared beyond its band gap and CMOS compatibility, and thus qualifies as an ideal material for transparent heaters enabling transmissive reconfigurable optics. Doped Si heaters have been employed to realize an array of waveguide-based PCM devices\cite{zhang2019miniature, zhang2019nonvolatile, zheng2020nonvolatile, erickson2023comparing} and recently a spatial light modulator prototype\cite{fang2023non}. In this work, we unveil the potential of an on-chip PCM-based metasurface that can reversibly change the optical transmissive behavior using an IR transparent silicon-on-insulator (SOI) microheater. This advancement opens the door to a variety of applications, such as compact, non-volatile, and non-mechanical tunable filters as is the focus of this study. Our experiments successfully demonstrate reversible switching of a non-mechanical optical filter using electrical pulses for almost 1250 cycles over an effective switching area of 120$\mu$m $\times$ 120$\mu$m. To the best of our knowledge, this marks the first demonstration of a large-area PCM-based metasurface with reliable switching across more than 1000 switching cycles. In addition, through a detailed device characterization, we identify the key factor that contributes to device failure and propose potential solutions to extend the metafilter operational lifetime beyond thousands of cycles.

\section{Methods} 
\subsection{Computational methods}

The electromagnetic simulations were performed using the rigorous coupled wave analysis (RCWA) MATLAB\textregistered\ library RETICOLO V8\cite{Hugonin2021RETICOLOAnalysis}. The refractive indices of Ge$_2$Sb$_2$Se$_4$Te (GSST) for intermediate crystallization states were interpolated based on the Maxwell-Garnett effective medium approximation\cite{Markel2016IntroductionTutorial} with the amorphous state as the host material and the crystalline phase as the inclusion using previously measured refractive indices of the two phases\cite{Zhang2019BroadbandPhotonics}. The simulations were performed from 0 to 16$^{\circ}$ incident angles in steps of 4$^{\circ}$ for both TE and TM-polarizations and averaged over the simulated incident angles and TE and TM polarizations. The field profile was simulated by assuming propagation from the silicon substrate into the meta-structure and the resulting values were multiplied by the transmittance from air into silicon averaged over wavelength, angle and polarizations. The thicknesses assumed in the simulations were as follows: silicon substrate, 1.01 $\mu$m buried oxide (BOX), 157 nm of doped silicon, 370 nm GSST, 310 nm  SiN$_x$, and air. Metafilters with various periods and hole sizes were fabricated for testing. For the Fourier transform infrared (FTIR) data collection and simulation, the dimensions were estimated as 445 nm $\times$ 445 nm size hole with 700 nm period. According to scanning electron microscopy (SEM) observation, a 60 nm layer of SiN$_x$ was coated on the bottom of the hole and on the walls of GSST resulting from the incomplete covering of the walls and the incomplete filling of the structure due to nonconformal sputtering deposition. 
%
\subsection{Fabrication Methods}

\begin{figure*}
\centering
{\includegraphics[scale= 0.58]{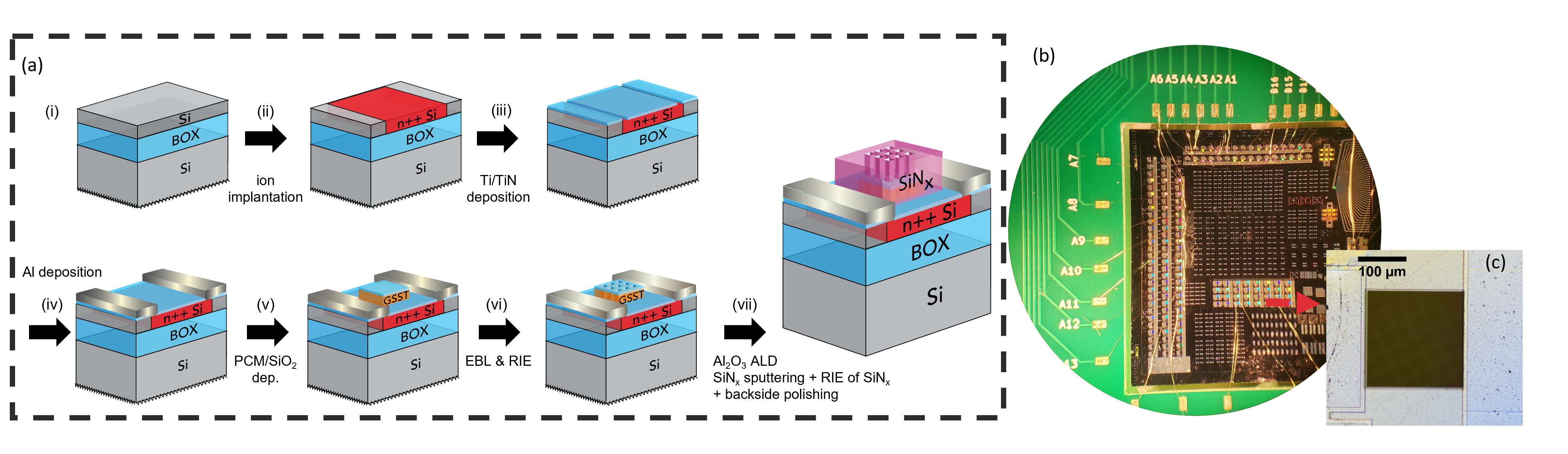}}
\caption{(a) Fabrication flow chart of the metasurface device. After encapsulation with silicon nitride and etch-back, aluminum metal apertures are deposited, and the sample is back-side polished via mechanical lapping to remove diffuse reflections. Afterwards, the sample is wire bonded to a printed circuit board. (b) Optical image of the chip after wirebonding. Different devices were fabricated with different periods, resulting in the various reflected colors. (c) Visible range optical micrograph of a fabricated metasurface device. The fringes observed in the images are Moir\'e artifacts from the specific magnification used, the camera pixel pitch and periodicity of the metasurface.
\label{fig_fabflowchart}}
\end{figure*}

The major steps in the fabrication process of the metasurface devices are illustrated in Fig. \ref{fig_fabflowchart}. Details of the SOI micro-heater fabrication were elaborated elsewhere \cite{rios2022ultra, popescu2023open}. The thickness of the buried oxide used was 1.01 $\mu$m and the doped Si layer thickness was 157 nm. Large squares for the initial blanket GSST deposition are defined in AZ nLOF 2020 using an MLA 150 Heidelberg at 375 nm. The prebaking and post-exposure baking were done at nominal 112 $^{\circ}$C for 60 s on a hot plate. An optimal dose was observed around or slightly above 155 mJ/cm$^{2}$. The development was done in AZ 300 MIF for around 70-80 s. Thin films of GSST of 370 nm thickness were deposited via thermal evaporation at a base pressure of 1.8 $\times$ 10$^{-6}$ Torr onto the chip. 10 nm of SiO$_2$ were evaporated via e-beam evaporation before resist removal for protection against oxidation. The resist was removed via overnight soaking in n-methylpyrrolidone (NMP), followed by acetone and  isopropanol (IPA) rinse. 50 nm of polymethyl methacrylate (PMMA) 495 K followed by 350 nm of ZEP 520 A e-beam resist were spin coated and pre-baked sequentially at 180 $^{\circ}C$ on a hot plate. The PMMA layer served the purpose of aiding resist removal after reactive ion etching (RIE). The metasurface mask was exposed at 50 kV using a current of 10 nA in an Elionix-HS50 with proximity effect correction at a dose of 170 $\mu$C/cm$^{2}$. The resist was developed in IPA: methyl isobutyl ketone (MIBK) 3:1 for 90 s and rinsed with IPA for 20 s. The resist was allowed to dry in air to prevent mechanical damage to the metasurface mask. The PCM was etched via RIE with CF$_4$:Ar (43:15 standard cubic centimeter per minute, sccm) at 0.5 Pa, 100 W bias power and 50 W inductively coupled plasma (ICP) power in a Samco-RIE 230iP for 230 s. The electron beam resist was then removed in NMP overnight, followed by the acetone and IPA rinse. Subsequently, the PCM metasurface was encapsulated with 20 nm of Al$_2$O$_3$ deposited at 150 $^{\circ}$C in a Cambridge Nanotech Savannah 200. A further protective layer of SiN$_x$ was deposited via reactive sputtering from a Si target and a Si$_3$N$_4$ target in N$_2$ : Ar atmosphere at 3 mTorr with 6:6 sccm flow rates using an AJA ATC Orion 5 chamber. The thickness of the deposited nitride was 310 nm. The region above the aluminum contacts was patterned and etched via RIE with a mixture of SF$_6$:Ar 6:20 sccm standard recipe. A 10 $\mu$m buffer region from the edge of the metal contacts was left unetched to prevent shorting in the subsequent steps of defining an aperture. Aluminum apertures are thermally evaporated around the heaters to limit the transmitted light just to the region with metasurfaces at a base pressure of around 9$\times$ 10$^{-6}$ Torr. To protect the structures during backside polishing, ~5 $\mu$m of PMMA 950 A were spin coated onto the substrate. The sample backside was then mechanically polished to enable transmissive measurements with minimal roughness scattering. The sample was secured to a support structure with CrystalBond\texttrademark \ wax before mechanical polishing with sequantially smaller particle size lapping paper (SiC 15 $\mu$m, 5 $\mu$m, 3 $\mu$m and Al$_2$O$_3$ 1 $\mu$m). The samples were removed from the support structure and rinsed in acetone and IPA to remove the wax and protective PMMA resist (\ref{fig_fabflowchart} (b,c)). Finally, the samples were mounted onto a custom-designed printed circuit board and wire bonded with a ball bonder MEI 1204D.

\subsection{Characterization Methods}
The optical response of our PCM-based filter was evaluated using a FTIR spectrometer with an external Mercury Cadmium Telluride (MCT) detector. The beam was focused onto the backside of the sample with a spot size of 250 $\mu$m in diameter using a reflective objective. After transmitting through the front of the sample, the beam was collected with another objective directed towards the detector. Since the focused beam is larger than the PCM metasurface, we deposited a 200 $\mu$m $\times$ 200 $\mu$m metallic aperture to restrict the beam so it only passes through the metasurface area. The transmission spectrum through a 200 $\mu$m $\times$ 200 $\mu$m square aperture (Thorlabs, Inc.), the same size as our metafilter, was also measured on the same setup to normalize the measured metasurface transmittance. In addition to the micro-FTIR measurement, a short-wave infrared (SWIR) camera (FLIR A6262) equipped with a long pass filter was used for detecting light within the range of 800 nm to 1700 nm. FTIR measurements allow us to capture the spectral transmission of the device, while the SWIR camera assists us in observing the devices' broadband response upon phase transformation.

\section{Results and Discussion}
In order to evaluate the performance of our PCM-based device upon switching between amorphous and crystalline states, we start our experiment by depositing a continuous patch of thin film GSST on top of the micro-heater prior to metasurface fabrication. The purpose of the test structure is to examine the contributing degradation mechanisms limiting endurance of PCM devices and inform mitigation strategies for our final metasurface device. The PCM film test structure has a slightly smaller coverage area (140 $\mu$m $\times$ 140 $\mu$m) than our subsequent metasurface device. The corresponding sheet resistance of the micro-heater is measured to be 42 $\Omega$/$\square$. For switching between the two states we used a 10 $\mu$s, 45V pulse for amorphization, and to switch back to crystalline state, a series of 200 ms priming pulses are sent to initiate nucleation area in the PCM and facilitate the phase transformation\cite{Orava2017Classical-nucleation-theoryMemory, aryana2023toward}. These priming pulses incrementally increase from 10 to 16 V and are applied to the sample before the crystallization pulse. Subsequently, after 1 sec from the priming pulses a 18 V crystallization pulse lasting for 1 second is applied. There are 30 sec pause between amorphization and crystallization to ensure complete thermal relaxation and that the device does not undergo electromigration.

\begin{figure*}
\centering
{\includegraphics[scale= 0.70]{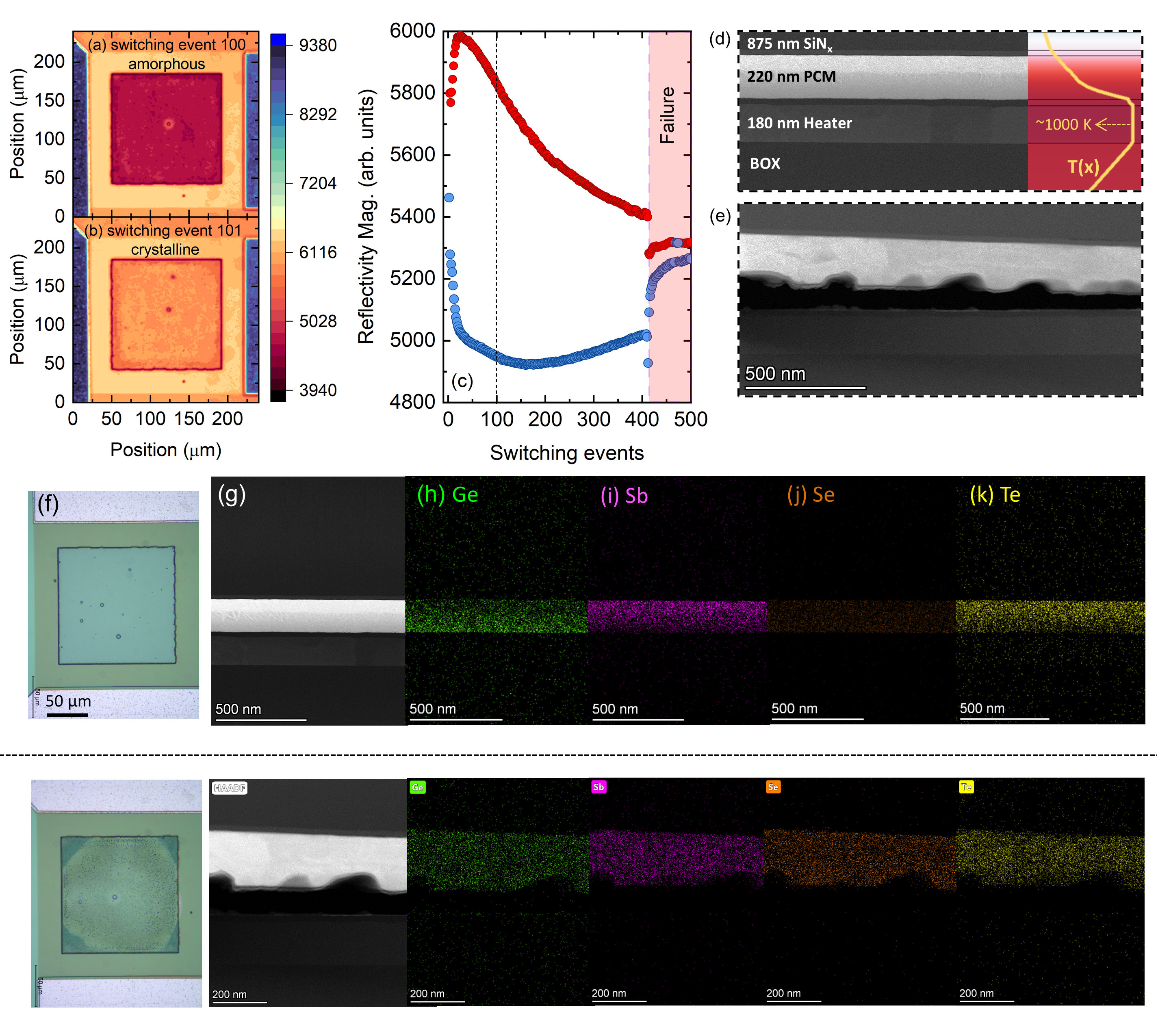}}
\caption{NIR image of an unpatterned PCM thin film test structure for degradation mechanism analysis, captured with front-side illumination during the switching events (a) 100 corresponding to amorphous state and (b) 101 corresponding to crystalline state. (c) The magnitude of reflected light from the surface of the PCM upon cycling detected by the IR camera. (d \& e) TEM micrographs of the device and its associated layers before and after cycling. (f) Optical image of pristine GSST thin film and (g) high-angle annular dark-field (HAADF) scanning transmission electron microscopy (STEM) micrograph of the film, along with (g-k) the elemental EDS maps of the thin film. (l) Optical image of a GSST thin film that was switched for 750 switching events and (m) its HAADF-STEM counterpart along with (n-r) the respective elemental maps from EDS.
\label{fig_X2}}
\end{figure*}

Figures \ref{fig_X2}(a,b) depict images captured by a SWIR camera upon transitioning from one phase to another, specifically at switching events of 100 and 101, corresponding to the amorphous and crystalline states, respectively. For better clarity, in this manuscript a ``switching event'' refers to a single phase transformation, while a ``cycle'' pertains to the process of transitioning to a different phase and subsequently returning to the original phase. Figure \ref{fig_X2}(c) shows the changes in surface reflectivity of the PCM upon cycling using the SWIR camera. As anticipated, the amorphous phase of GSST exhibits higher transparency in the near-infrared spectrum compared to the crystalline phase. Consequently, the reflection from the PCM-covered area is less pronounced in the amorphous phase compared to the crystalline phase. As can be seen in Fig. \ref{fig_X2}(c), at the initial stages of the cycling, the device shows changes in the reflectivity. As cycling continues, the switching contrast gradually enhances after several tens of cycles. Based on our examination through EDS microscopy Fig.\ref{fig_X2}(f-r) and line-scans (refer to Supplementary Note 3), we attribute the burn-in behavior of the device\cite{2013xiongSelfAligned} to the presence of Ge-rich regions in close proximity to the heater, and Sb and Te rich regions away from the heater, which are a consequence of the deposition process. The Ge-rich area is expected to have a higher crystallization and melting temperature, which in turn gives rise to non-uniform switching temperatures throughout the thickness of the GSST layer. 
With repeated cycling, we observe the migration of Ge away from the heater. As hundreds of cycles are completed, we begin to see small areas within the PCM-covered region stay unchanged upon switching, which gradually expand in size, which accounts for the gradual decrease of reflectance contrast from 20 to 400 switching events. A comprehensive set of images documenting the switching process from start to finish has been compiled into a video. This video is available in Supplementary Movie 1 and offers visual insight into the formation of damage on the micro-heater surface during the switching period.

We evaluate the failure mechanism by performing transmission electron microscopy (TEM) before and after 375 cycles (750 switching events) as well as using computational models to estimate the temperature distribution in the device. Figure \ref{fig_X2}(d) presents the layer configuration in the device and the model used in our simulations, along with the corresponding temperature increase at the heater. Our observations reveal that during cycling, the temperature peaks at the electrode/PCM interface and reaches as high as 1000 K during amorphization cycle, giving rise to maximum thermal stress at the interface. Interestingly, our TEM result after failure, given in Fig. \ref{fig_X2}(e), indicates that the device undergoes delamination from this very interface. Multiple factors could contribute to the delamination of the PCM, such as volume contraction of GSST upon crystallization (approximately 3-5\% \cite{shalaginov2021reconfigurable,Aryana2021SuppressedGe2Sb2Se4Te}), disparate thermal expansion coefficients between the layers, inadequate adhesion between the PCM and the electrode, and insufficient thermo-mechanical durability of the encapsulating SiN$_x$ layer. Our prior research has demonstrated that a SiN$_x$ layer with inferior mechanical properties can significantly impair the device's lifetime\cite{popescu2023long}. 


\begin{figure*}[h!]
\centering
{\includegraphics[scale= 0.8]{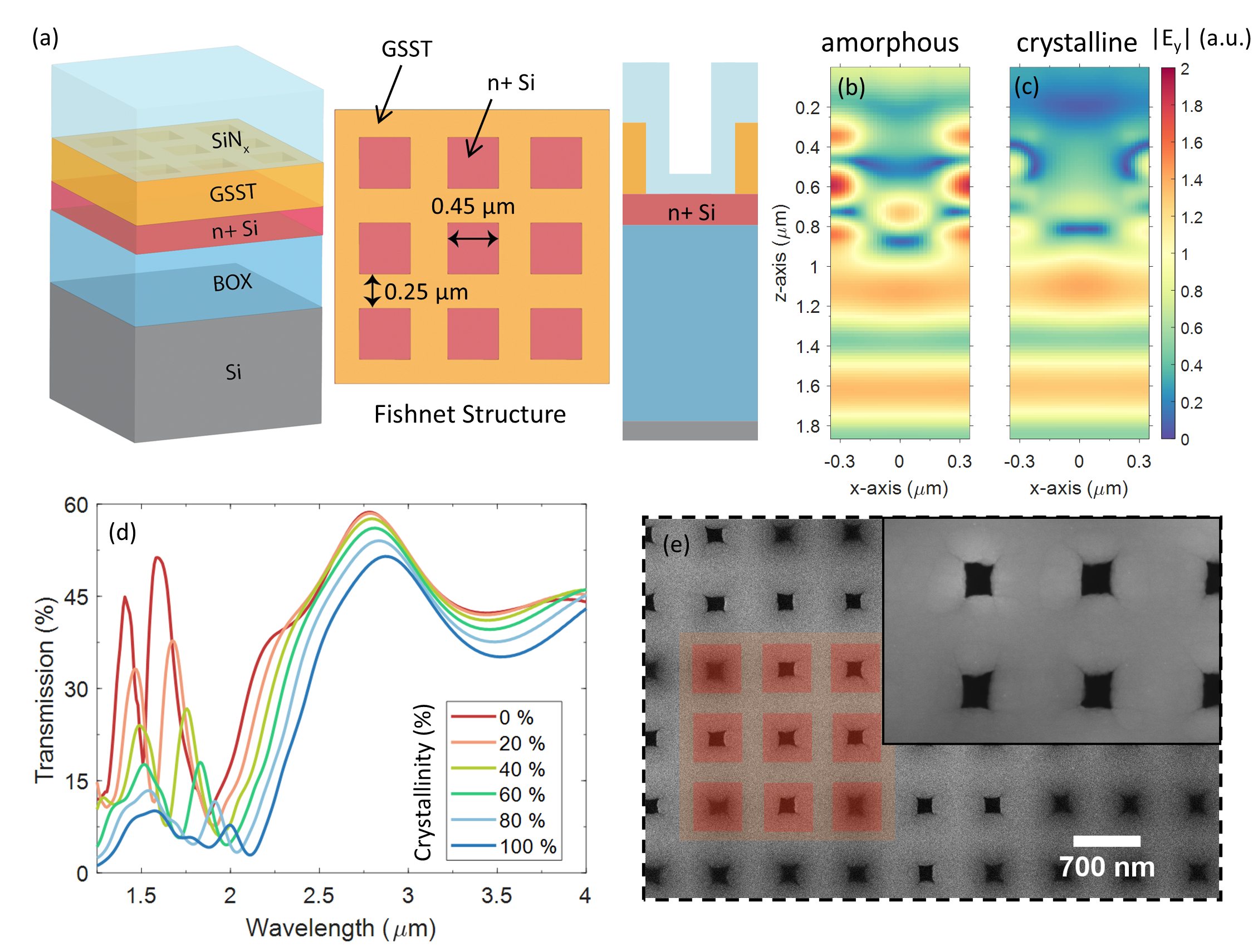}}
\caption{Simplified device structure assumed in RCWA simulations (a) along with $|E_y|$ in the meta-atom in the fully amorphous (b) and crystalline (c) state at 1420 nm for normal incidence, polarization along $\hat{Y}$. Simulated transmission curves averaged over TE and TM and over angles up to 16$^{\circ}$. An SEM image of the device. The inset shows the air holes remaining after silicon nitride deposition along with the morphology changes at the edges, likely due to raised nitride due to a higher deposition rate at the GSST edges (e).  
\label{fig_simulations}}
\end{figure*}

Evaluating the thin-film PCM on the microheater provided us with crucial insights into the switching behavior and failure mechanisms in this specific device architecture. This understanding, derived from the performance analysis of the heater with thin-film PCM, is vital in informing the design process for a more effective and durable filter. Since delamination constitute the main degradation mechanism, we hypothesize that a fishnet metasurface structure as illustrated in Fig. \ref{fig_simulations} (a,e), can alleviate the mechanical failure and yields durable filter devices. First, the top capping layers (Al$_2$O$_3$ + SiN$_x$) that encapsulates the PCM can be in direct contact with the SiO$_2$ covered heater surface through holes in the fishnet structure. Given the excellent adhesion between SiO$_2$ and atomic layer deposited Al$_2$O$_3$, we expect that the contacts between them can act as anchor points to impede delamination of the PCM structures \cite{ding2011influence,takakura2023room}. Second, the large surface areas of the fishnet structures can also effectively relieve the mechanical stress. Furthermore, compared to typical ``pillar'' metasurfaces, we found that the fishnet design improved switching consistency since crystallization can efficiently propagate across the entire structure through crystalline growth without necessitating nucleation, which is highly stochastic in small structures exemplified by ``pillar'' meta-atoms.

Based on the fishnet geometry, we deployed RCWA simulations to engineer a PCM-based metasurface specifically for filtering purposes. The simulated transmission spectra of the sample as a function of crystalline volume fraction are given in Fig. \ref{fig_simulations} (d). The main features are a significant decrease in transmission centered at 1.6 $\mu$m wavelength with increased crystalline fraction, along with a red-shift of the spectral peaks observed due to an increase in the index of GSST. The transmission envelope overall shrinks due to increased absorption in the crystalline state. Simulations of the propagating electric field through the structure show higher field intensity in the PCM region and nitride cladding in the amorphous state for a wavelength with higher transmission (Fig. \ref{fig_simulations} b). The increased absorption and index of the crystalline structure leads to a weaker field amplitude in the PCM region, with most of the signal being reflected back (Fig. \ref{fig_simulations} c). 

\begin{figure*}[h]
\centering
{\includegraphics[scale= 0.74]{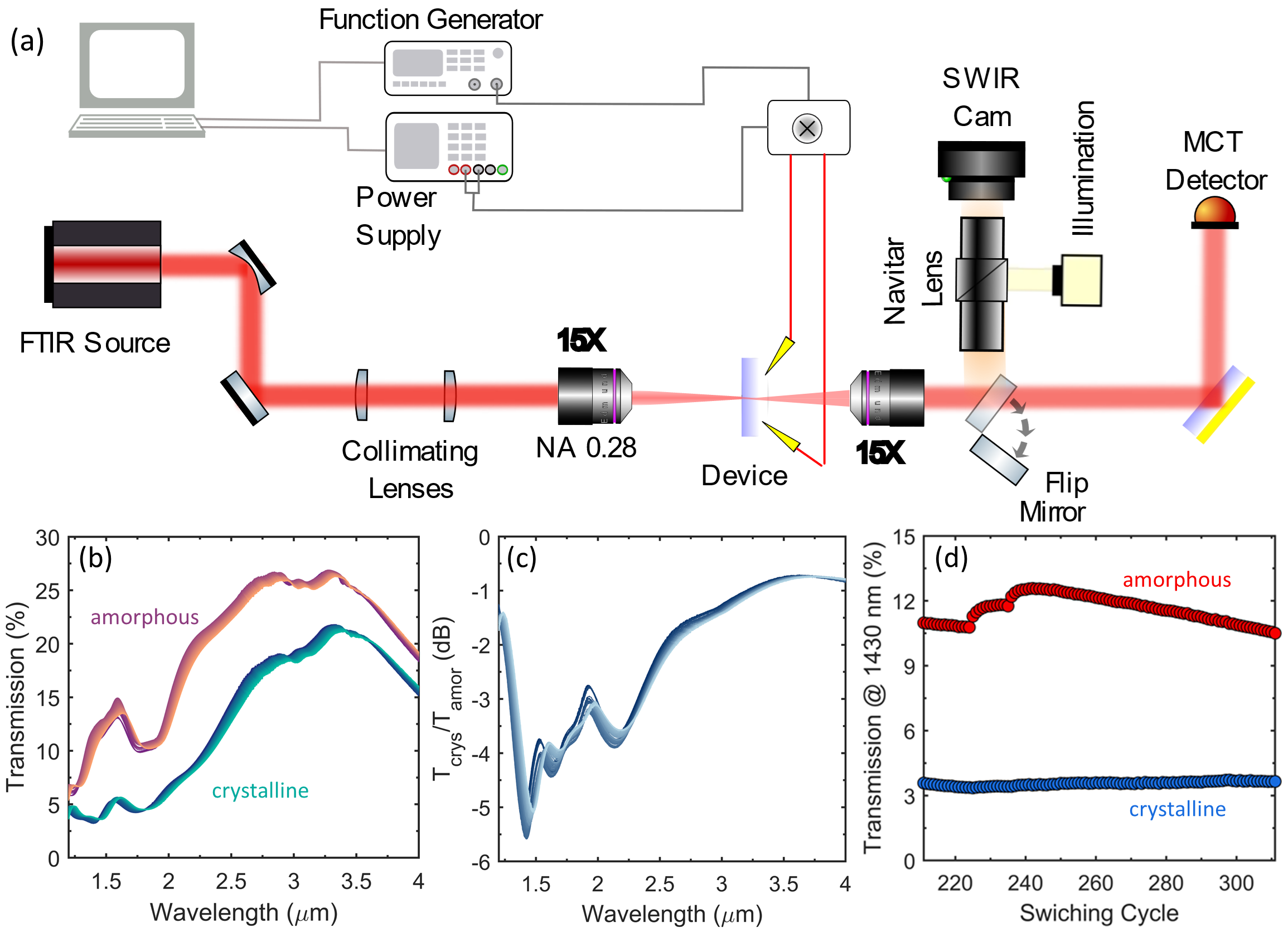}}
\caption{(a) Schematic of the measurement setup for the FTIR spectral data collection. (b) 
Transmission spectra upon switching between amorphous and crystalline states over the course of 100 cycles. (c) Transmission ratio between amorphous and crystalline phases as a function of wavelength. The inset shows the drift of the maximum contrast to larger wavelengths over the 100 cycles. (d) Transmission magnitude for amorphous and crystalline state at 1430 nm. The jump points were triggered by slight adjustments to the amorphization voltage.
\label{fig_FTIR}}
\end{figure*}

Following the fabrication of the PCM-based metasurface on the microheater, we proceeded to assess the device's transmission characteristics across a wavelength range spanning from 1.25 to 4 $\mu$m using FTIR spectroscopy. Given the device's compact dimensions, we developed an in-house micro-FTIR setup. In Fig. \ref{fig_FTIR}(a), one can observe a schematic of the experimental setup used for this purpose. As depicted in Fig. \ref{fig_FTIR}(a), the output from the FTIR spectrometer, which is approximately 2 inches in diameter, passes through a beam condenser comprising a pair of CaF$_2$ lenses for resizing it to fit the aperture of the focusing objective, which focuses the beam to match the metasurface size. After the beam passes through the device, the light is recollected with a secondary objective and directed towards an external Mercury Cadmium Telluride (MCT) detector. A flip mirror is inserted on the path of the beam to align the beam to the specific device under test. For a more detailed explanation of the measurement procedure, including signal strength and background collection, refer to Supplementary Note 1.

The experimental results presented in Fig. \ref{fig_FTIR}(b) show trends similar to those predicted in simulations, in particular suppression of the transmission peak at 1.6 $\mu$m wavelength as well as the overall transmission reduction at wavelengths above 2.0 $\mu$m during the transition from the amorphous to the crystalline state. Peak transmission contrast exceeding 5.5 dB was obtained near the 1.43 $\mu$m wavelength (Fig. \ref{fig_FTIR} d). To validate switching consistency, the measured spectra over 100 switching cycles were overlaid in Fig. \ref{fig_FTIR} (c) and (d). To provide further insight into these results, the absolute transmittance at 1430 nm as a function of cycle number for both amorphous and crystalline states are plotted in Fig. \ref{fig_FTIR} (d). The switching behavior was observed to be relatively stable throughout the 100 cycles. The kinks at cycles 225 and 236 for the amorphous state were due to adjustments to the amorphization voltage as part of the switching parameter optimization process.

\begin{figure*}[h!]
\centering
{\includegraphics[scale= 0.5]{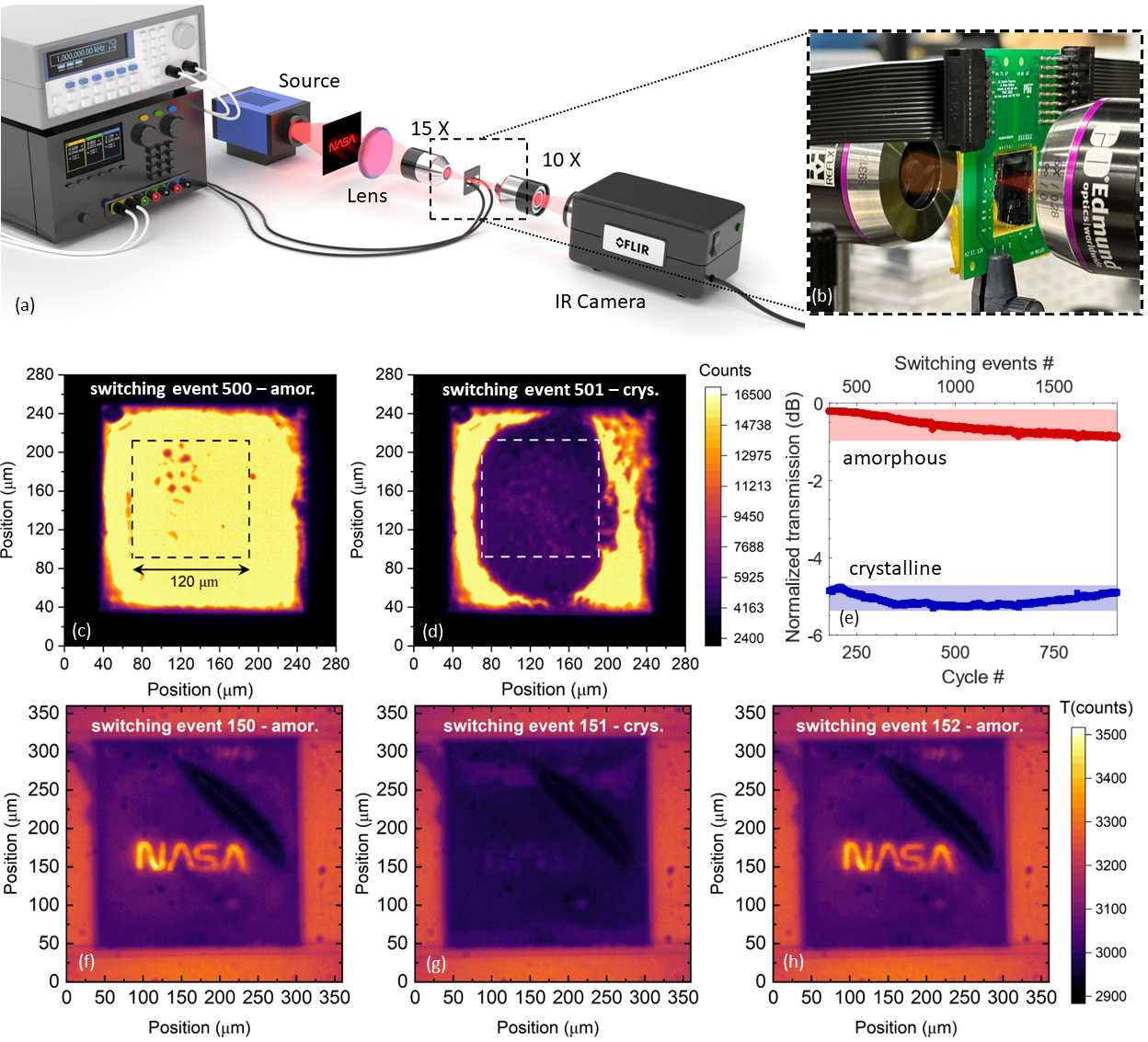}}
\caption{(a) An illustration of the measurement arrangement, featuring a focused beam passing through the filter. (b) A real-time image of the sample being tested. The infrared camera image captured when the broadband beam is passing through the device in its (c) amorphous and (d) crystalline phase. (e) The mean count value across 14000 pixels within the square marked by dashed lines, over 700 stable cycles. The full testing data is available in Supplementary Note 1 Fig. S3 (b).
Panels (f-h) illustrate the imaging performance of a PCM-based filter in its activated (on) and deactivated (off) states. The dark streak over NASA worm logo is a delaminated region.
\label{fig_X4_camera}}
\end{figure*}

To further evaluate the performance of the fabricated filters and gain insights into the switching characteristics across the heater's surface, we employed a SWIR camera to image the entire surface of the PCM-based metasurface upon switching. For this, similar to FTIR measurements as depicted in Fig. \ref{fig_X4_camera}(a), we directed the broadband IR beam onto the 200$\mu$m $\times$ 200$\mu$m device, and used the camera as the detector. After each switching event, the SWIR camera captures an image of the device surface and monitors any potential damage during cycling. In this arrangement, the sample is backlit by a focused IR beam, while the camera, positioned in front, records the transmitted light intensity. Figure \ref{fig_X4_camera}(b) displays a real-time image of the sample being tested, highlighting the focused beam as it passes through the PCM-based tunable filter. A complete video detailing the switching cycles of this filter from start to finish can be found in Supplementary Movie 2.

A further demonstration of the filter's capabilities is depicted in panels Fig. \ref{fig_X4_camera}(f-h). In this setup, a glass slide with an etched metal mask featuring the NASA worm logo was positioned in the beam path immediately after the light source. This mask selectively allows light to pass only through the etched NASA logo as illustrated in Fig. \ref{fig_X4_camera}(a). Through this arrangement, we can turn the NASA logo on and off by changing the phase of the PCM in the metasurface using electrical pulses. As seen in panel Fig. \ref{fig_X4_camera}(f), in the amorphous state, the filter is in high transmission mode and thus, the NASA logo is clearly visible at the camera, whereas in the crystalline state when the filter is in blocking mode, the intensity of the logo is significantly suppressed. Similarly, as depicted in Fig. \ref{fig_X4_camera}(h), applying an amorphization pulse, can effectively restore the NASA logo. This process can be done reversibly leveraging just electrical stimuli. A video featuring the images from each cycle is available in Supplementary Movie 3 for further reference.

In this study, we designed, fabricated, and demonstrated the performance of an all solid-state, non-volatile reconfigurable filter without the use of any mechanical moving parts. Our work takes advantage of a PCM-based metasurface that is capable of switching between a high- and low-transmittance state by applying electrical pulses within a microelectronic circuitry framework. We have also identified delamination as the leading failure mechanisms in the device and implemented a fishnet metasurface design to enhance its endurance to 1250 cycles, which represents a 25-fold improvement over previously reported PCM-based metasurfaces of comparable aperture size. This work introduces a reliable platform for switching PCM-based metasurfaces, opening up new possibilities in low-power reconfigurable optics and programmable photonics across various applications.


\section{Disclaimer}
Specific vendor and manufacturer names are explicitly mentioned only to accurately describe the test hardware. The use of vendor and manufacturer names does not imply an endorsement by the U.S. Government nor does it imply that the specified equipment is the best available.

\bibliography{references.bib}

\begin{thebibliography}{10}
\urlstyle{rm}
\expandafter\ifx\csname url\endcsname\relax
  \def\url#1{\texttt{#1}}\fi
\expandafter\ifx\csname urlprefix\endcsname\relax\def\urlprefix{URL }\fi
\expandafter\ifx\csname doiprefix\endcsname\relax\def\doiprefix{DOI: }\fi
\providecommand{\bibinfo}[2]{#2}
\providecommand{\eprint}[2][]{\url{#2}}

\bibitem{horst2023tbit}
\bibinfo{author}{Horst, Y.} \emph{et~al.}
\newblock \bibinfo{journal}{\bibinfo{title}{Tbit/s line-rate satellite feeder links enabled by coherent modulation and full-adaptive optics}}.
\newblock {\emph{\JournalTitle{Light: Science \& Applications}}} \textbf{\bibinfo{volume}{12}}, \bibinfo{pages}{153} (\bibinfo{year}{2023}).

\bibitem{hampson2021adaptive}
\bibinfo{author}{Hampson, K.~M.} \emph{et~al.}
\newblock \bibinfo{journal}{\bibinfo{title}{Adaptive optics for high-resolution imaging}}.
\newblock {\emph{\JournalTitle{Nature Reviews Methods Primers}}} \textbf{\bibinfo{volume}{1}}, \bibinfo{pages}{68} (\bibinfo{year}{2021}).

\bibitem{davies2012adaptive}
\bibinfo{author}{Davies, R.} \& \bibinfo{author}{Kasper, M.}
\newblock \bibinfo{journal}{\bibinfo{title}{Adaptive optics for astronomy}}.
\newblock {\emph{\JournalTitle{Annual Review of Astronomy and Astrophysics}}} \textbf{\bibinfo{volume}{50}}, \bibinfo{pages}{305--351} (\bibinfo{year}{2012}).

\bibitem{salter2019adaptive}
\bibinfo{author}{Salter, P.~S.} \& \bibinfo{author}{Booth, M.~J.}
\newblock \bibinfo{journal}{\bibinfo{title}{Adaptive optics in laser processing}}.
\newblock {\emph{\JournalTitle{Light: Science \& Applications}}} \textbf{\bibinfo{volume}{8}}, \bibinfo{pages}{110} (\bibinfo{year}{2019}).

\bibitem{algorri2019recent}
\bibinfo{author}{Algorri, J.~F.}, \bibinfo{author}{Zografopoulos, D.~C.}, \bibinfo{author}{Urruchi, V.} \& \bibinfo{author}{S{\'a}nchez-Pena, J.~M.}
\newblock \bibinfo{journal}{\bibinfo{title}{Recent advances in adaptive liquid crystal lenses}}.
\newblock {\emph{\JournalTitle{Crystals}}} \textbf{\bibinfo{volume}{9}}, \bibinfo{pages}{272} (\bibinfo{year}{2019}).

\bibitem{madec2012overview}
\bibinfo{author}{Madec, P.-Y.}
\newblock \bibinfo{title}{Overview of deformable mirror technologies for adaptive optics and astronomy}.
\newblock In \emph{\bibinfo{booktitle}{Adaptive Optics Systems III}}, vol. \bibinfo{volume}{8447}, \bibinfo{pages}{22--39} (\bibinfo{organization}{SPIE}, \bibinfo{year}{2012}).

\bibitem{beichman2012science}
\bibinfo{author}{Beichman, C.~A.} \emph{et~al.}
\newblock \bibinfo{title}{Science opportunities with the near-ir camera ({NIRCam}) on the james webb space telescope ({JWST})}.
\newblock In \emph{\bibinfo{booktitle}{Space Telescopes and Instrumentation 2012: Optical, Infrared, and Millimeter Wave}}, vol. \bibinfo{volume}{8442}, \bibinfo{pages}{973--983} (\bibinfo{organization}{SPIE}, \bibinfo{year}{2012}).

\bibitem{ren2015tailoring}
\bibinfo{author}{Ren, Y.-X.}, \bibinfo{author}{Lu, R.-D.} \& \bibinfo{author}{Gong, L.}
\newblock \bibinfo{journal}{\bibinfo{title}{Tailoring light with a digital micromirror device}}.
\newblock {\emph{\JournalTitle{Annalen der physik}}} \textbf{\bibinfo{volume}{527}}, \bibinfo{pages}{447--470} (\bibinfo{year}{2015}).

\bibitem{gyger2021reconfigurable}
\bibinfo{author}{Gyger, S.} \emph{et~al.}
\newblock \bibinfo{journal}{\bibinfo{title}{Reconfigurable photonics with on-chip single-photon detectors}}.
\newblock {\emph{\JournalTitle{Nature communications}}} \textbf{\bibinfo{volume}{12}}, \bibinfo{pages}{1408} (\bibinfo{year}{2021}).

\bibitem{gu2023reconfigurable}
\bibinfo{author}{Gu, T.}, \bibinfo{author}{Kim, H.~J.}, \bibinfo{author}{Rivero-Baleine, C.} \& \bibinfo{author}{Hu, J.}
\newblock \bibinfo{journal}{\bibinfo{title}{Reconfigurable metasurfaces towards commercial success}}.
\newblock {\emph{\JournalTitle{Nature Photonics}}} \textbf{\bibinfo{volume}{17}}, \bibinfo{pages}{48--58} (\bibinfo{year}{2023}).

\bibitem{zhang2021electrically}
\bibinfo{author}{Zhang, Y.} \emph{et~al.}
\newblock \bibinfo{journal}{\bibinfo{title}{Electrically reconfigurable non-volatile metasurface using low-loss optical phase-change material}}.
\newblock {\emph{\JournalTitle{Nature Nanotechnology}}} \textbf{\bibinfo{volume}{16}}, \bibinfo{pages}{661--666} (\bibinfo{year}{2021}).

\bibitem{abdollahramezani2022electrically}
\bibinfo{author}{Abdollahramezani, S.} \emph{et~al.}
\newblock \bibinfo{journal}{\bibinfo{title}{Electrically driven reprogrammable phase-change metasurface reaching 80\% efficiency}}.
\newblock {\emph{\JournalTitle{Nature communications}}} \textbf{\bibinfo{volume}{13}}, \bibinfo{pages}{1696} (\bibinfo{year}{2022}).

\bibitem{Wu2019DynamicMetasurfaces}
\bibinfo{author}{Wu, P.~C.} \emph{et~al.}
\newblock \bibinfo{journal}{\bibinfo{title}{{Dynamic beam steering with all-dielectric electro-optic III–V multiple-quantum-well metasurfaces}}}.
\newblock {\emph{\JournalTitle{Nature Communications}}} \textbf{\bibinfo{volume}{10}}, \bibinfo{pages}{3654}, \doiprefix\url{10.1038/s41467-019-11598-8} (\bibinfo{year}{2019}).

\bibitem{Iyer2018UniformMetalenses}
\bibinfo{author}{Iyer, P.~P.}, \bibinfo{author}{Decrescent, R.~A.}, \bibinfo{author}{Lewi, T.}, \bibinfo{author}{Antonellis, N.} \& \bibinfo{author}{Schuller, J.~A.}
\newblock \bibinfo{journal}{\bibinfo{title}{{Uniform Thermo-Optic Tunability of Dielectric Metalenses}}}.
\newblock {\emph{\JournalTitle{Physical Review Applied}}} \textbf{\bibinfo{volume}{10}}, \bibinfo{pages}{044029}, \doiprefix\url{10.1103/PhysRevApplied.10.044029} (\bibinfo{year}{2018}).

\bibitem{Kim2019PhaseMetasurfaces}
\bibinfo{author}{Kim, Y.} \emph{et~al.}
\newblock \bibinfo{journal}{\bibinfo{title}{{Phase Modulation with Electrically Tunable Vanadium Dioxide Phase-Change Metasurfaces}}}.
\newblock {\emph{\JournalTitle{Nano Letters}}} \textbf{\bibinfo{volume}{19}}, \bibinfo{pages}{3961--3968}, \doiprefix\url{10.1021/acs.nanolett.9b01246} (\bibinfo{year}{2019}).

\bibitem{Yin2017BeamMetasurfaces}
\bibinfo{author}{Yin, X.} \emph{et~al.}
\newblock \bibinfo{journal}{\bibinfo{title}{{Beam switching and bifocal zoom lensing using active plasmonic metasurfaces}}}.
\newblock {\emph{\JournalTitle{Light: Science \& Applications}}} \textbf{\bibinfo{volume}{6}}, \bibinfo{pages}{e17016}, \doiprefix\url{10.1038/lsa.2017.16} (\bibinfo{year}{2017}).

\bibitem{julian2020reversible}
\bibinfo{author}{Julian, M.~N.}, \bibinfo{author}{Williams, C.}, \bibinfo{author}{Borg, S.}, \bibinfo{author}{Bartram, S.} \& \bibinfo{author}{Kim, H.~J.}
\newblock \bibinfo{journal}{\bibinfo{title}{Reversible optical tuning of \uppercase{G}e\uppercase{S}b\uppercase{T}e phase-change metasurface spectral filters for mid-wave infrared imaging}}.
\newblock {\emph{\JournalTitle{Optica}}} \textbf{\bibinfo{volume}{7}}, \bibinfo{pages}{746--754} (\bibinfo{year}{2020}).

\bibitem{raeis2017metasurfaces}
\bibinfo{author}{Raeis-Hosseini, N.} \& \bibinfo{author}{Rho, J.}
\newblock \bibinfo{journal}{\bibinfo{title}{Metasurfaces based on phase-change material as a reconfigurable platform for multifunctional devices}}.
\newblock {\emph{\JournalTitle{Materials}}} \textbf{\bibinfo{volume}{10}}, \bibinfo{pages}{1046} (\bibinfo{year}{2017}).

\bibitem{de2020reconfigurable}
\bibinfo{author}{de~Galarreta, C.~R.} \emph{et~al.}
\newblock \bibinfo{journal}{\bibinfo{title}{Reconfigurable multilevel control of hybrid all-dielectric phase-change metasurfaces}}.
\newblock {\emph{\JournalTitle{Optica}}} \textbf{\bibinfo{volume}{7}}, \bibinfo{pages}{476--484} (\bibinfo{year}{2020}).

\bibitem{wang2023varifocal}
\bibinfo{author}{Wang, M.} \emph{et~al.}
\newblock \bibinfo{journal}{\bibinfo{title}{Varifocal metalens using tunable and ultralow-loss dielectrics}}.
\newblock {\emph{\JournalTitle{Advanced Science}}} \textbf{\bibinfo{volume}{10}}, \bibinfo{pages}{2204899} (\bibinfo{year}{2023}).

\bibitem{shaltout2019spatiotemporal}
\bibinfo{author}{Shaltout, A.~M.}, \bibinfo{author}{Shalaev, V.~M.} \& \bibinfo{author}{Brongersma, M.~L.}
\newblock \bibinfo{journal}{\bibinfo{title}{Spatiotemporal light control with active metasurfaces}}.
\newblock {\emph{\JournalTitle{Science}}} \textbf{\bibinfo{volume}{364}}, \bibinfo{pages}{eaat3100} (\bibinfo{year}{2019}).

\bibitem{wuttig2017phase}
\bibinfo{author}{Wuttig, M.}, \bibinfo{author}{Bhaskaran, H.} \& \bibinfo{author}{Taubner, T.}
\newblock \bibinfo{journal}{\bibinfo{title}{Phase-change materials for non-volatile photonic applications}}.
\newblock {\emph{\JournalTitle{Nature photonics}}} \textbf{\bibinfo{volume}{11}}, \bibinfo{pages}{465--476} (\bibinfo{year}{2017}).

\bibitem{zhang2021myths}
\bibinfo{author}{Zhang, Y.} \emph{et~al.}
\newblock \bibinfo{journal}{\bibinfo{title}{Myths and truths about optical phase change materials: A perspective}}.
\newblock {\emph{\JournalTitle{Applied Physics Letters}}} \textbf{\bibinfo{volume}{118}} (\bibinfo{year}{2021}).

\bibitem{zhang2019miniature}
\bibinfo{author}{Zhang, H.} \emph{et~al.}
\newblock \bibinfo{journal}{\bibinfo{title}{Miniature multilevel optical memristive switch using phase change material}}.
\newblock {\emph{\JournalTitle{ACS Photonics}}} \textbf{\bibinfo{volume}{6}}, \bibinfo{pages}{2205--2212} (\bibinfo{year}{2019}).

\bibitem{zhang2019nonvolatile}
\bibinfo{author}{Zhang, H.} \emph{et~al.}
\newblock \bibinfo{journal}{\bibinfo{title}{Nonvolatile waveguide transmission tuning with electrically-driven ultra-small {GST} phase-change material}}.
\newblock {\emph{\JournalTitle{Science Bulletin}}} \textbf{\bibinfo{volume}{64}}, \bibinfo{pages}{782--789} (\bibinfo{year}{2019}).

\bibitem{zheng2020nonvolatile}
\bibinfo{author}{Zheng, J.} \emph{et~al.}
\newblock \bibinfo{journal}{\bibinfo{title}{Nonvolatile electrically reconfigurable integrated photonic switch enabled by a silicon {PIN} diode heater}}.
\newblock {\emph{\JournalTitle{Advanced Materials}}} \textbf{\bibinfo{volume}{32}}, \bibinfo{pages}{2001218} (\bibinfo{year}{2020}).

\bibitem{erickson2023comparing}
\bibinfo{author}{Erickson, J.~R.} \emph{et~al.}
\newblock \bibinfo{journal}{\bibinfo{title}{Comparing the thermal performance and endurance of resistive and pin silicon microheaters for phase-change photonic applications}}.
\newblock {\emph{\JournalTitle{Optical Materials Express}}} \textbf{\bibinfo{volume}{13}}, \bibinfo{pages}{1677--1688} (\bibinfo{year}{2023}).

\bibitem{fang2023non}
\bibinfo{author}{Fang, Z.} \emph{et~al.}
\newblock \bibinfo{journal}{\bibinfo{title}{Non-volatile phase-only transmissive spatial light modulators}}.
\newblock {\emph{\JournalTitle{arXiv preprint arXiv:2307.12103}}}  (\bibinfo{year}{2023}).

\bibitem{Hugonin2021RETICOLOAnalysis}
\bibinfo{author}{Hugonin, J.~P.} \& \bibinfo{author}{Lalanne, P.}
\newblock \bibinfo{title}{{RETICOLO software for grating analysis}} (\bibinfo{year}{2021}).

\bibitem{Markel2016IntroductionTutorial}
\bibinfo{author}{Markel, V.~A.}
\newblock \bibinfo{journal}{\bibinfo{title}{{Introduction to the Maxwell Garnett approximation: tutorial}}}.
\newblock {\emph{\JournalTitle{Journal of the Optical Society of America A}}} \textbf{\bibinfo{volume}{33}}, \doiprefix\url{10.1364/josaa.33.001244} (\bibinfo{year}{2016}).

\bibitem{Zhang2019BroadbandPhotonics}
\bibinfo{author}{Zhang, Y.} \emph{et~al.}
\newblock \bibinfo{journal}{\bibinfo{title}{{Broadband transparent optical phase change materials for high-performance nonvolatile photonics}}}.
\newblock {\emph{\JournalTitle{Nature Communications}}} \textbf{\bibinfo{volume}{10}}, \bibinfo{pages}{4279}, \doiprefix\url{10.1038/s41467-019-12196-4} (\bibinfo{year}{2019}).

\bibitem{rios2022ultra}
\bibinfo{author}{R{\'\i}os, C.} \emph{et~al.}
\newblock \bibinfo{journal}{\bibinfo{title}{Ultra-compact nonvolatile phase shifter based on electrically reprogrammable transparent phase change materials}}.
\newblock {\emph{\JournalTitle{PhotoniX}}} \textbf{\bibinfo{volume}{3}}, \bibinfo{pages}{26} (\bibinfo{year}{2022}).

\bibitem{popescu2023open}
\bibinfo{author}{Popescu, C.-C.} \emph{et~al.}
\newblock \bibinfo{journal}{\bibinfo{title}{An open-source multifunctional testing platform for optical phase change materials}}.
\newblock {\emph{\JournalTitle{Small Science}}} \textbf{\bibinfo{volume}{n/a}}, \bibinfo{pages}{2300098}, \doiprefix\url{https://doi.org/10.1002/smsc.202300098} (\bibinfo{year}{2023}).
\newblock \eprint{https://onlinelibrary.wiley.com/doi/pdf/10.1002/smsc.202300098}.

\bibitem{Orava2017Classical-nucleation-theoryMemory}
\bibinfo{author}{Orava, J.} \& \bibinfo{author}{Greer, A.~L.}
\newblock \bibinfo{journal}{\bibinfo{title}{{Classical-nucleation-theory analysis of priming in chalcogenide phase-change memory}}}.
\newblock {\emph{\JournalTitle{Acta Materialia}}} \textbf{\bibinfo{volume}{139}}, \bibinfo{pages}{226--235}, \doiprefix\url{10.1016/j.actamat.2017.08.013} (\bibinfo{year}{2017}).

\bibitem{aryana2023toward}
\bibinfo{author}{Aryana, K.} \emph{et~al.}
\newblock \bibinfo{journal}{\bibinfo{title}{Toward accurate thermal modeling of phase change material based photonic devices}}.
\newblock {\emph{\JournalTitle{Small}}}  (\bibinfo{year}{2023}).

\bibitem{2013xiongSelfAligned}
\bibinfo{author}{Xiong, F.} \emph{et~al.}
\newblock \bibinfo{journal}{\bibinfo{title}{Self-aligned nanotube–nanowire phase change memory}}.
\newblock {\emph{\JournalTitle{Nano Letters}}} \textbf{\bibinfo{volume}{13}}, \bibinfo{pages}{464--469}, \doiprefix\url{10.1021/nl3038097} (\bibinfo{year}{2013}).
\newblock \bibinfo{note}{PMID: 23259592}, \eprint{https://doi.org/10.1021/nl3038097}.

\bibitem{shalaginov2021reconfigurable}
\bibinfo{author}{Shalaginov, M.~Y.} \emph{et~al.}
\newblock \bibinfo{journal}{\bibinfo{title}{Reconfigurable all-dielectric metalens with diffraction-limited performance}}.
\newblock {\emph{\JournalTitle{Nature communications}}} \textbf{\bibinfo{volume}{12}}, \bibinfo{pages}{1225} (\bibinfo{year}{2021}).

\bibitem{Aryana2021SuppressedGe2Sb2Se4Te}
\bibinfo{author}{Aryana, K.} \emph{et~al.}
\newblock \bibinfo{journal}{\bibinfo{title}{{Suppressed electronic contribution in thermal conductivity of Ge2Sb2Se4Te}}}.
\newblock {\emph{\JournalTitle{Nature Communications}}} \textbf{\bibinfo{volume}{12}}, \bibinfo{pages}{1--9}, \doiprefix\url{10.1038/s41467-021-27121-x} (\bibinfo{year}{2021}).

\bibitem{popescu2023long}
\bibinfo{author}{Popescu, C.-C.} \emph{et~al.}
\newblock \bibinfo{title}{Long live {O-PCM}s: Understanding reliability challenges of optical phase change materials}.
\newblock In \emph{\bibinfo{booktitle}{CLEO: Science and Innovations}}, \bibinfo{pages}{STh1O--5} (\bibinfo{organization}{Optica Publishing Group}, \bibinfo{year}{2023}).

\bibitem{ding2011influence}
\bibinfo{author}{Ding, J.} \emph{et~al.}
\newblock \bibinfo{journal}{\bibinfo{title}{The influence of substrate on the adhesion behaviors of atomic layer deposited aluminum oxide films}}.
\newblock {\emph{\JournalTitle{Surface and Coatings Technology}}} \textbf{\bibinfo{volume}{205}}, \bibinfo{pages}{2846--2851} (\bibinfo{year}{2011}).

\bibitem{takakura2023room}
\bibinfo{author}{Takakura, R.}, \bibinfo{author}{Murakami, S.}, \bibinfo{author}{Watanabe, K.} \& \bibinfo{author}{Takigawa, R.}
\newblock \bibinfo{journal}{\bibinfo{title}{Room-temperature bonding of {Al}$_2${O}$_3$ thin films deposited using atomic layer deposition}}.
\newblock {\emph{\JournalTitle{Scientific reports}}} \textbf{\bibinfo{volume}{13}}, \bibinfo{pages}{3581} (\bibinfo{year}{2023}).

\end{thebibliography}

\section*{Data availability.} 
    The data that support the findings of this study are available from the corresponding author upon reasonable request.

\section*{Acknowledgements}
    The authors appreciate the support by Mr. Ronald Neale in graphic design. This work was carried out in part through the use of MIT.nano's facilities. This work was partially supported by the Air Force Office of Scientific Research (AFOSR), award numbers FA9550-22-1-0456 and FA9550-22-1-0532. This research was sponsored by the National Aeronautics and Space Administration (NASA) through a contract with ORAU. The views and conclusions contained in this document are those of the authors and should not be interpreted as representing the official policies, either expressed or implied, of the National Aeronautics and Space Administration (NASA) or the U.S. Government. The U.S. Government is authorized to reproduce and distribute reprints for Government purposes notwithstanding any copyright notation herein.

\section*{Author contributions}
    K.A., C.C.P., H.J.K., and J.H. designed the study. C.C.P. performed the photonic device design and fabricated the metasurface samples. H.J.K. and P.G. fabricated the logos. K.A., P.G., and C.C.P. performed the characterization. K.A. and K.P.D. performed the COMSOL simulations. S.V. and V.L. fabricated the SOI micro-heater platform. H.B.B. and T.W.L. performed STEM sample preparation and imaging. C.A.R.O. and Y.Z. designed the SOI micro-heater platform. K.A., C.C.P. and J.H. wrote the manuscript. M.K. and K.A.R. provided the bulk phase change materials. All authors contributed to technical discussions and revising the manuscript.
    
\section*{Competing interests}   
    The authors declare no competing interests.
    
\noindent \textbf{Correspondence} and requests for materials should be addressed to K.A. or H.J.K.

\end{document}


\date{}


\begin{center}
 \textbf{(Supplementary Information)} 
\end{center}

\clearpage

\section*{Supplementary Note 1}
\begin{figure*}[h!]
\centering
{\includegraphics[scale= 0.75]{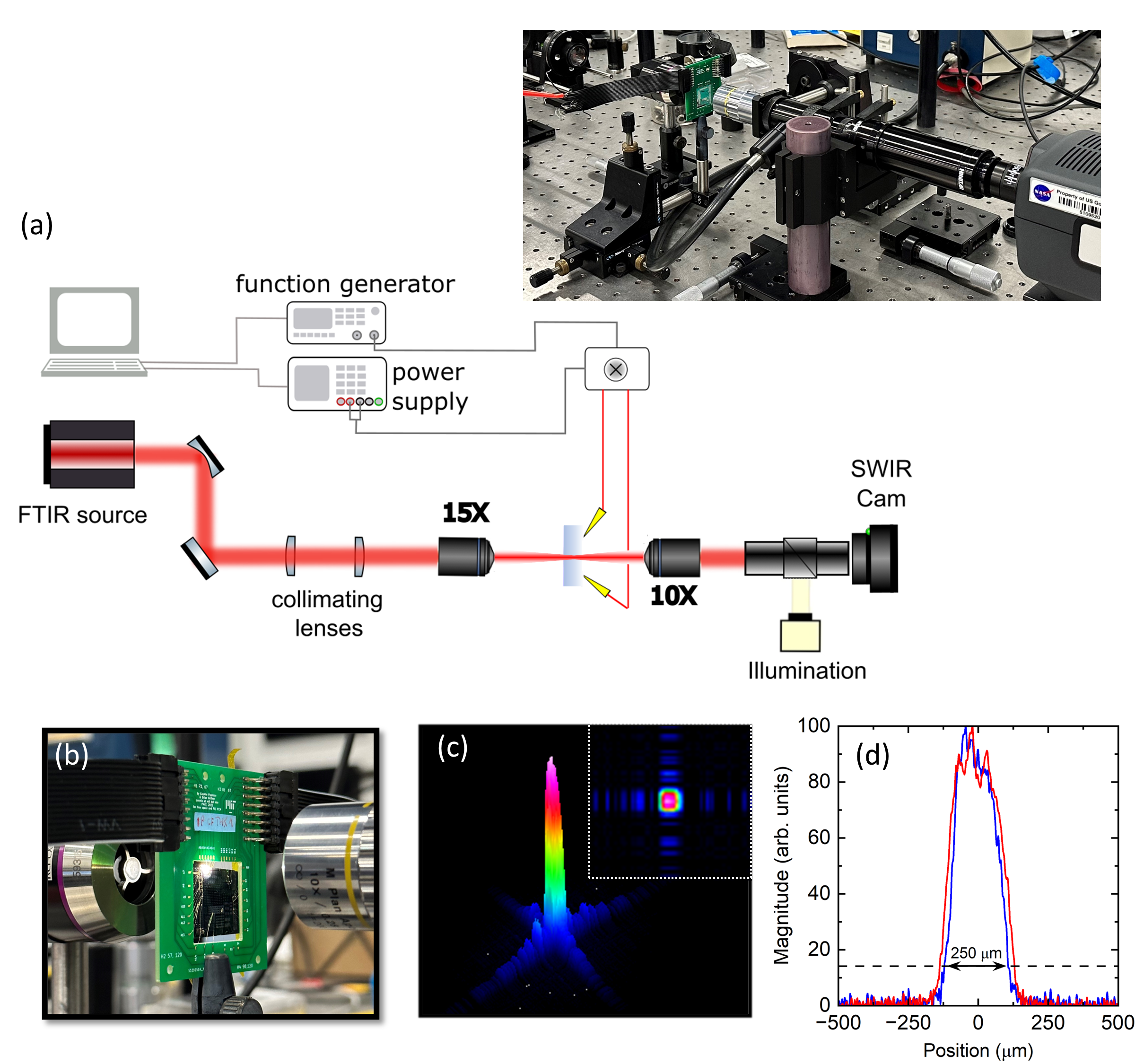}}
\caption{(a) Near IR optical transmission configuration for imaging the device and optical image of the measuring setup 
(b) device under test, (c) 3D beam shape from scanning slit beam profiler, (d) beam size in X and Y direction.
\label{fig_SchemBeamSize}}
\end{figure*}

\textbf{Micro-FTIR measurements.} To evaluate the performance of our PCM-based reconfigurable filters, we employ Fourier transform infrared (FTIR) spectroscopy with an external Mercury Cadmium Telluride (MCT) detector. We use parabolic mirrors, CaF$_2$ lenses, and focusing objectives to condense and focus the beam to sub-millimeter length scale to probe individual devices. 
The setup for measuring the transmission of the filter is depicted in Fig. \ref{fig_SchemBeamSize}(a). This micro-FTIR setup is capable of effectively reducing the broadband IR beam, from around 2 inches in diameter at the output of the FTIR source, to nearly 250 $\mu$m at the sample's surface. To mitigate issues such as chromatic aberration and electromagnetic spectrum loss due to optical absorption by the focusing components in conventional refractive objectives, we utilize 15X reflective objectives that are designed for broad spectral range with a transmission exceeding 94\% across the 1000-11000 nm spectrum.

\begin{figure*}[h!]
\centering
{\includegraphics[scale= 0.7]{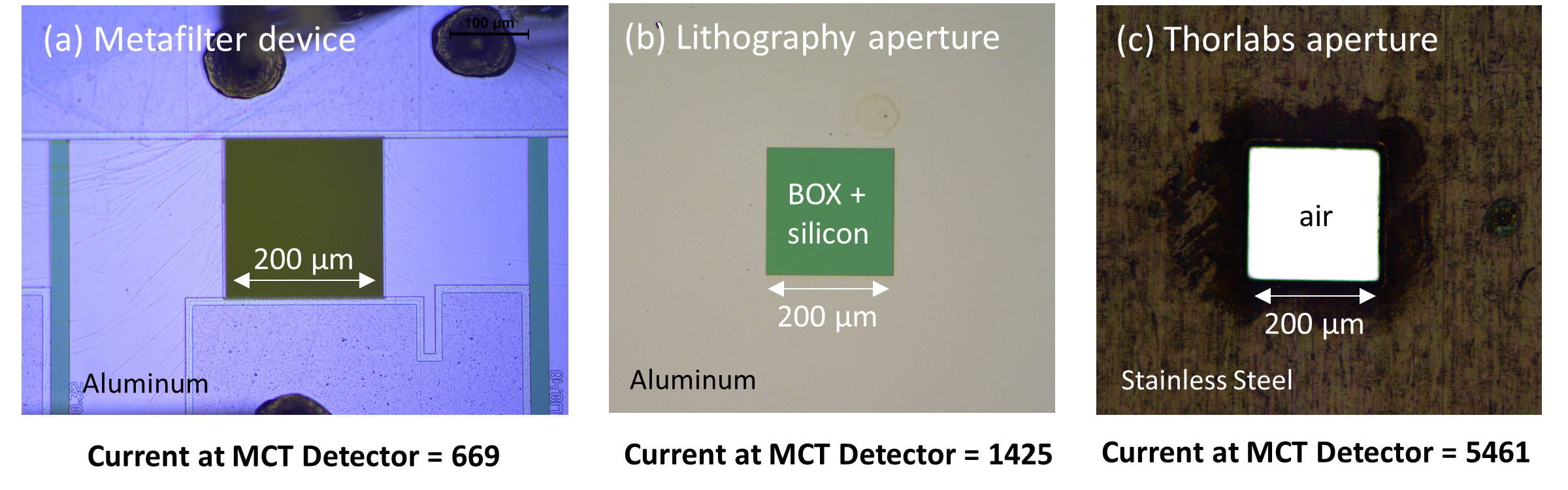}}
\caption{(a)  Metafilter filter with aluminum aperture and electrode contacts for electrical bias \& (b) 200 $\mu$m aperture deposited on the same chip as the metafilter for collecting background for FTIR measurements (BOX + Si + protective cladding). (c) 200 $\mu$m square aperture purchased from Thorlabs.
\label{fig_aperture}}
\end{figure*}

Figure \ref{fig_SchemBeamSize}(b) shows the device under test, which is positioned at the focal point of the two objectives. For improved image clarity, we swap the reflective objective on the camera side with a conventional refractive objective. We measure the spot size using a dual scanning slit beam profiler (Thorlabs BP209-IR) that operates in the wavelength range of 900-2500 nm. The results, depicted in Fig. \ref{fig_SchemBeamSize}(c,d), include a qualitative 3D and 2D representation of the beam, along with a beam intensity graph. This graph illustrates the beam's size in the X and Y directions, revealing a beam diameter of 250 $\mu$m using the 1/e$^2$ method, and a nearly Gaussian profile. Given that the beam's size marginally exceeds the area of the phase change material (PCM), the background collection for FTIR would not be accurate as part of the beam is blocked by the electrodes. For this, we restrict our testing to a 200 $\mu$m by 200 $\mu$m square microheater. This is done by depositing aluminum around the heater to define the test region precisely. A 10 $\mu$m buffer region of SiN$_x$ is left un-etched on the contact pads in order to prevent shorting of the device (region visible in  Fig. \ref{fig_aperture} (a)). This arrangement allows us to exclusively measure the light passing through the square microheater, effectively isolating it for accurate FTIR analysis.

We used a square aperture from Thorlabs (S200QK), which enabled us to collect background in air. We recorded the current at the MCT detector in three scenarios: (i) with the metafilter, (ii) with the chip aperture, and (iii) with the Thorlabs aperture (Fig. \ref{fig_aperture}). The maximum signal detected at the MCT detector for the beam alone, without any aperture, was 6100 counts. Adding the Thorlabs square aperture reduced this value to 5461 counts, a reasonable decrease given the beam's larger diameter (250 micrometers) compared to the aperture. However, for the aperture deposited on the chip, the signal drops to 1425 counts. This large drop in the signal compared to the aperture in the air could be due to reflection losses at the air-silicon interface or translation of the focus point when inserting the substrate. The addition of the microheater and the PCM metasurface and their interlayers drop the signal compared to the aperture on the chip by almost a factor of two. The FTIR measurements reported in the main manuscript are taken with respect to air as the background using the Thorlabs square aperture.

\begin{figure*}[h!]
\centering
{\includegraphics[scale= 0.75]{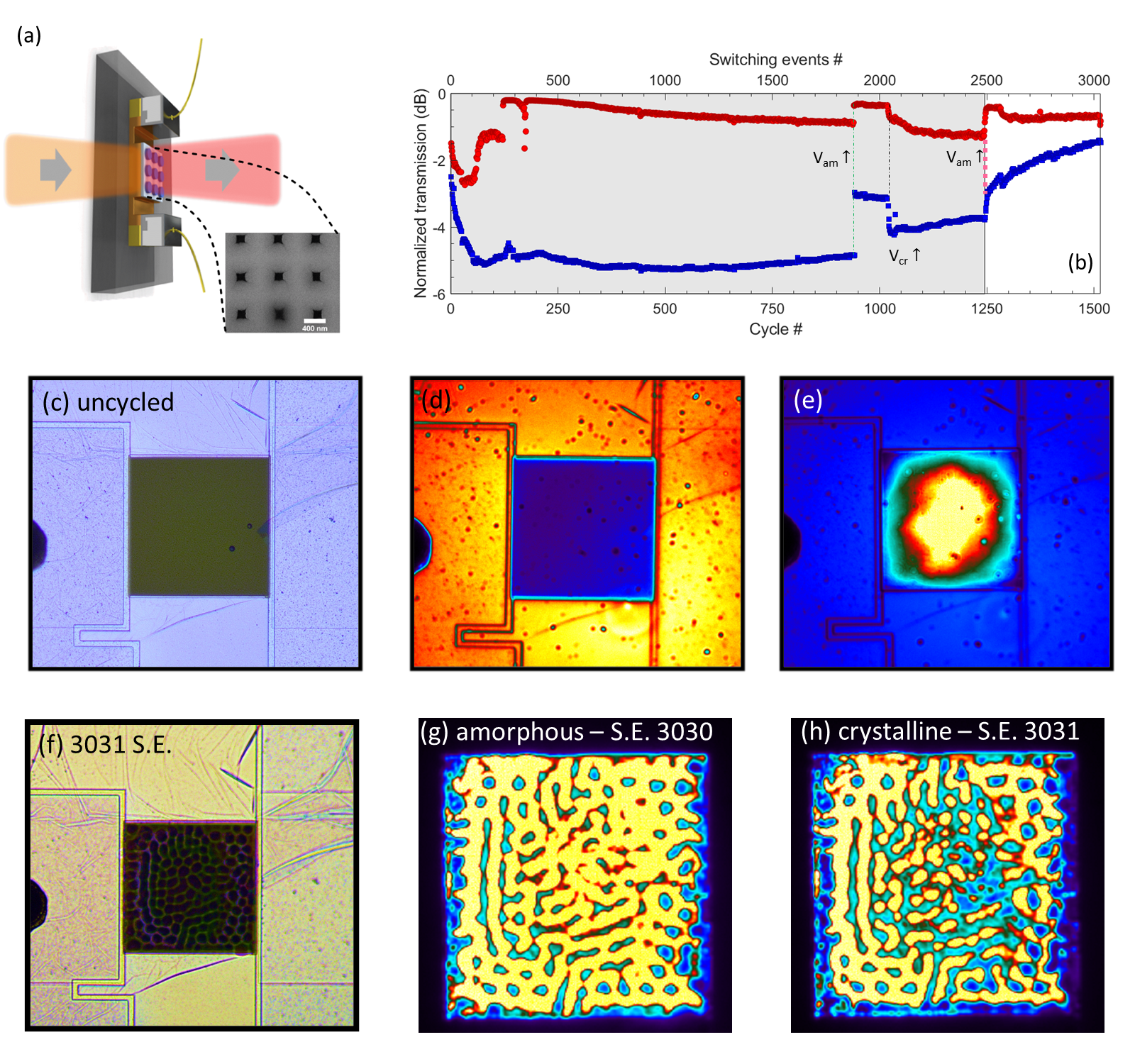}}
\caption{(a) Schematic of the configuration for transmission measurements of the metafilter. (b) 
Normalized transmission based on maximum transmitted light through the metafilter upon cycling. Several notable events in the device testing are highlighted by the dashed lines and nearby  V$_{am} \uparrow$ and V$_{cr} \uparrow$, meaning the increase in the voltage for the amorphization and crystallization pulse, respectively(c) Image of 200 $\mu$m $\times$ 200 $\mu$m metafilter on a microheater using a visible camera, (d) image of the same metafilter illuminated from the front without the focused IR beam using a SWIR camera, (e) image of the PCM-based filter with focused black-body beam on the device. (f) Image of the metafilter after 3031 switching events under visible microscope. (g) NIR image of the metafilter in amorphous state and (h) crystalline state after 3030 and 3031 switching events, respectively. 
\label{fig_Cam_EndDev}}
\end{figure*}

Fig. \ref{fig_Cam_EndDev}(a) illustrates a schematic of the device, highlighting the focused beam traversing the metafilter. The color shift in the diagram symbolizes the filter's ability to selectively block certain wavelengths. Fig. \ref{fig_Cam_EndDev}(b) presents the absolute value as a function of switching events (S.E.), corresponding to the data in  Fig. 6 (e) of the main manuscript. 
This graph depicts the filter's performance over 3000 switching events. To prevent overheating and potential damage to the sample, we carefully increased the voltage to enhance the contrast between the amorphous and crystalline states. The contrast peaks at the 400\textsuperscript{th} switching event /200\textsuperscript{th} cycle and maintains stability for the next 1400 events /700 cycles. Around the 1900th switching event, we raised the amorphization voltage to regain the initial high transmission levels. Although this led to a more complete amorphization and improved transmission, it also inadvertently increased transmission in the crystalline state, indicating less material crystallizing at the same voltage. Consequently, after several hundred cycles, we increased the crystallization voltage. This, however, resulted in a decrease in the amplitude of amorphization. Further increasing the amorphization voltage pushed the device towards its damage threshold, and soon after, we noticed signs of damage and a diminishing contrast in the image upon cycling. To highlight the difference in transmission between the amorphous and crystalline states, we normalized the data in Fig. \ref{fig_Cam_EndDev}(b) relative to the highest count value recorded by the camera, which is 16383 (14-bit depth). The normalized transmission values are then expressed in decibels (dB) in Fig. \ref{fig_Cam_EndDev}(c). The camera count data was collected from a central 120 $\mu$m $\times$ 120 $\mu$m region of interest that showed stability throughout the course of testing, as some level of delamination was observed incipient from around the edges of the device. 

Fig. \ref{fig_Cam_EndDev}(c\&f) show the same device which had experienced a little over 3000 switching events before and after the experiment. Fig. \ref{fig_Cam_EndDev}(d\&e), obtained using a a short-wave infrared (SWIR) camera, shows the metafilter in two scenarios: when the IR beam is blocked and when it is unblocked, allowing it to pass through the device. In Fig. \ref{fig_Cam_EndDev}(d), we present an image of the device captured by the SWIR camera prior to the cycling with front illumination, while the IR beam is blocked. When the beam is unblocked, allowing it to pass through the metafilter, nearly the entire area of the metafilter becomes illuminated, as demonstrated in Fig. \ref{fig_Cam_EndDev}(e). Prior to initiating the cycling for this experiment, the front illumination was switched off to focus exclusively on the impact of light traversing the metafilter. Similarly, we also present images captured by the camera after the device had completed 3000 cycles, showing both in the amorphous and crystalline phases as the IR beam passes through the metafilter Fig. \ref{fig_Cam_EndDev}(g,h) .

\section*{Supplementary Note 2}
\begin{figure*}[h!]
\centering
{\includegraphics[scale= 0.5]{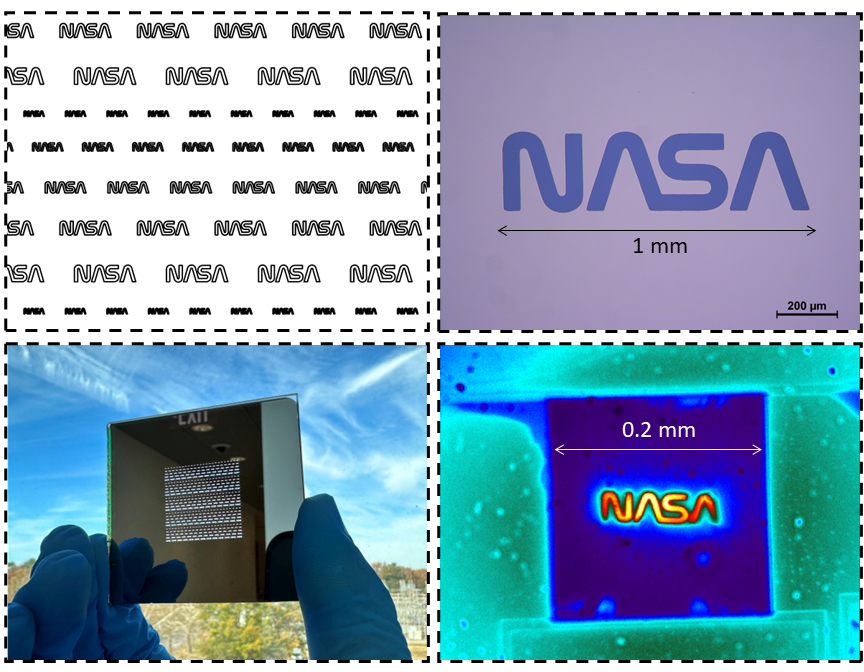}}
\caption{(a) Designed and (b,c) fabricated logo after laser lithography. (d) transmitted IR light through the logo after passing through the metafilter captured by the SWIR camera.
\label{fig_logo}}
\end{figure*}

\textbf{Laser lithography - NASA logo}
The NASA logos (DXF, Drawing Exchange Format) were patterned using Cr (chrome) on a soda lime glass substrate by using a maskless laser lithography system (DWL 66+ model manufactured by Heidelberg-instruments) with a 10mm write head (0.7 $\mu$m resolution). The DXF file was transferred to LIC format (Laser Internal Code) to be read by the DWL 66+ machine. The soda lime glass substrate (2.54 cm $\times$ 2.54 cm), with 200 nm thickness Chrome film and 1518 5K photo resist was prepared for the patterning of the logos. The logos were exposed films, the mask was developed using AZ400K 1:4 PR developer (1 part of concentrate and 4 parts of deionization-water, AZ electronic materials as the manufacturer) for 1 minute, Cr etchant 1020 (Transene Company, Inc.) for 1 minute, and AZ400k developer (no diluted as stripper, AZ electronic materials) for 3 minutes at room temperature to remove all residue. The targeted design layout, a resulting individual logo on the glass substrate and the whole fabricated glass chip are visible in Fig. \ref{fig_logo} (a), (b) and (c) respectively. By inserting the mask in the path of the beam before it is focused, the image is projected only on the metasurface device (Fig. \ref{fig_logo} (d)).

\section*{Supplementary Note 3}
\textbf{Scanning transmission electron microscopy (STEM) sample preparation and imaging.}

Samples for STEM and TEM observation before cycling in the as-deposited state and after 375 cycles were prepared by using a lift-out technique with ion-beam milling in a focused ion beam system (Helios G5 UX, Thermo Fisher Scientific). A protective amorphous carbon layer was applied over the region of interest before ion-milling. To minimize the sidewall damage and sufficiently thin the specimen for electron transparency, final ion-milling was carried out at a low voltage of $\sim$2 kV. All TEM/STEM images were obtained with a Cs-corrected transmission electron microscope (Titan cubed G2 60-300, Thermo Fisher Scientific) at 300 kV. Chemical composition analysis with EDS was carried out in the Titan cubed G2 at 300 kV along with four integrated silicon-drift EDS detectors at a collection solid angle of 0.7 srad. The probe current was about 100 pA with a scanning time of < 300sec.

\begin{figure*}
\centering
{\includegraphics[scale= 0.5]{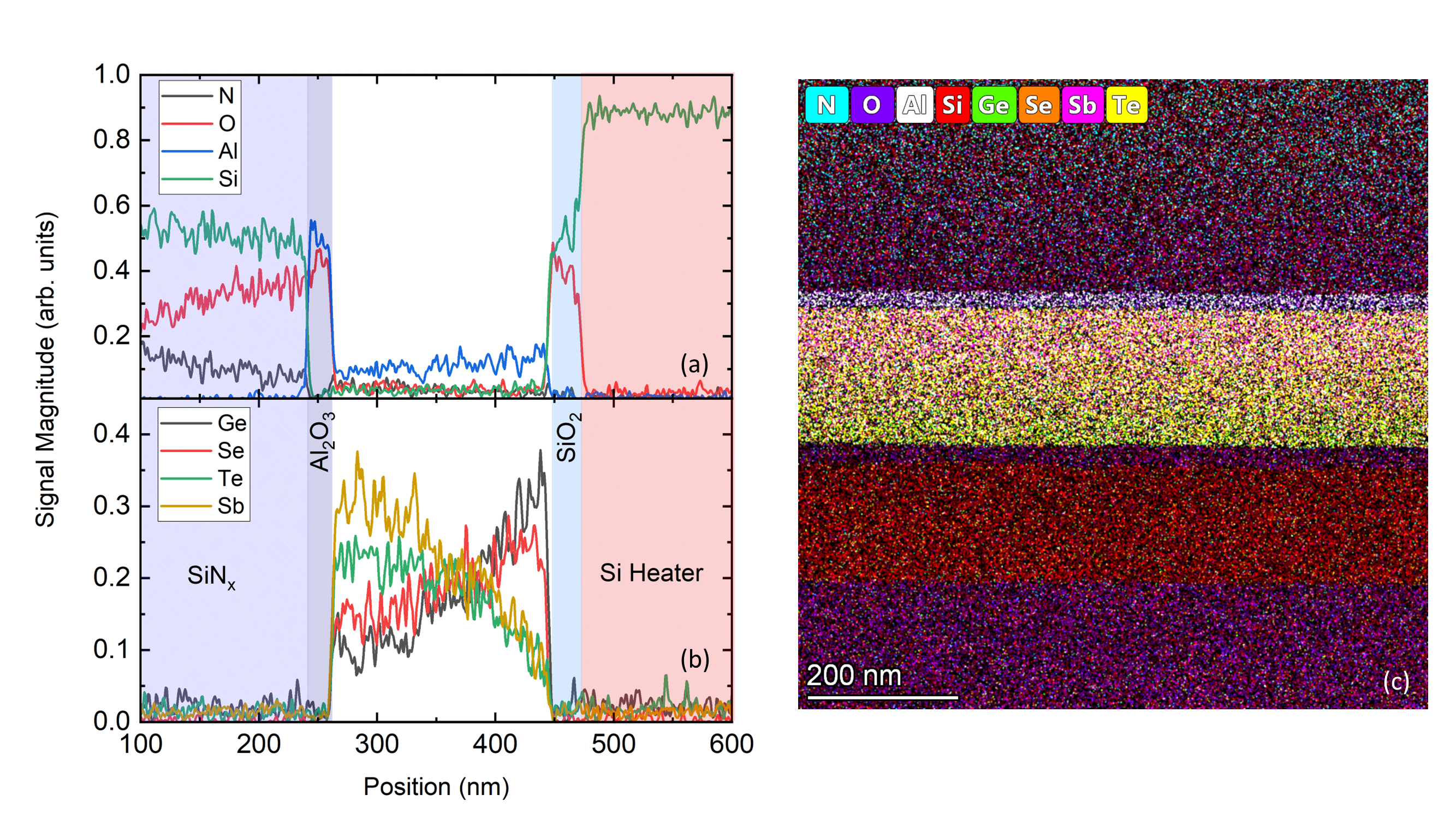}}
\caption{Layers composition in the device before cycling in the as-deposited state.
\label{fig_EDS_01}}
\end{figure*}

\begin{figure*}
\centering
{\includegraphics[scale= 0.5]{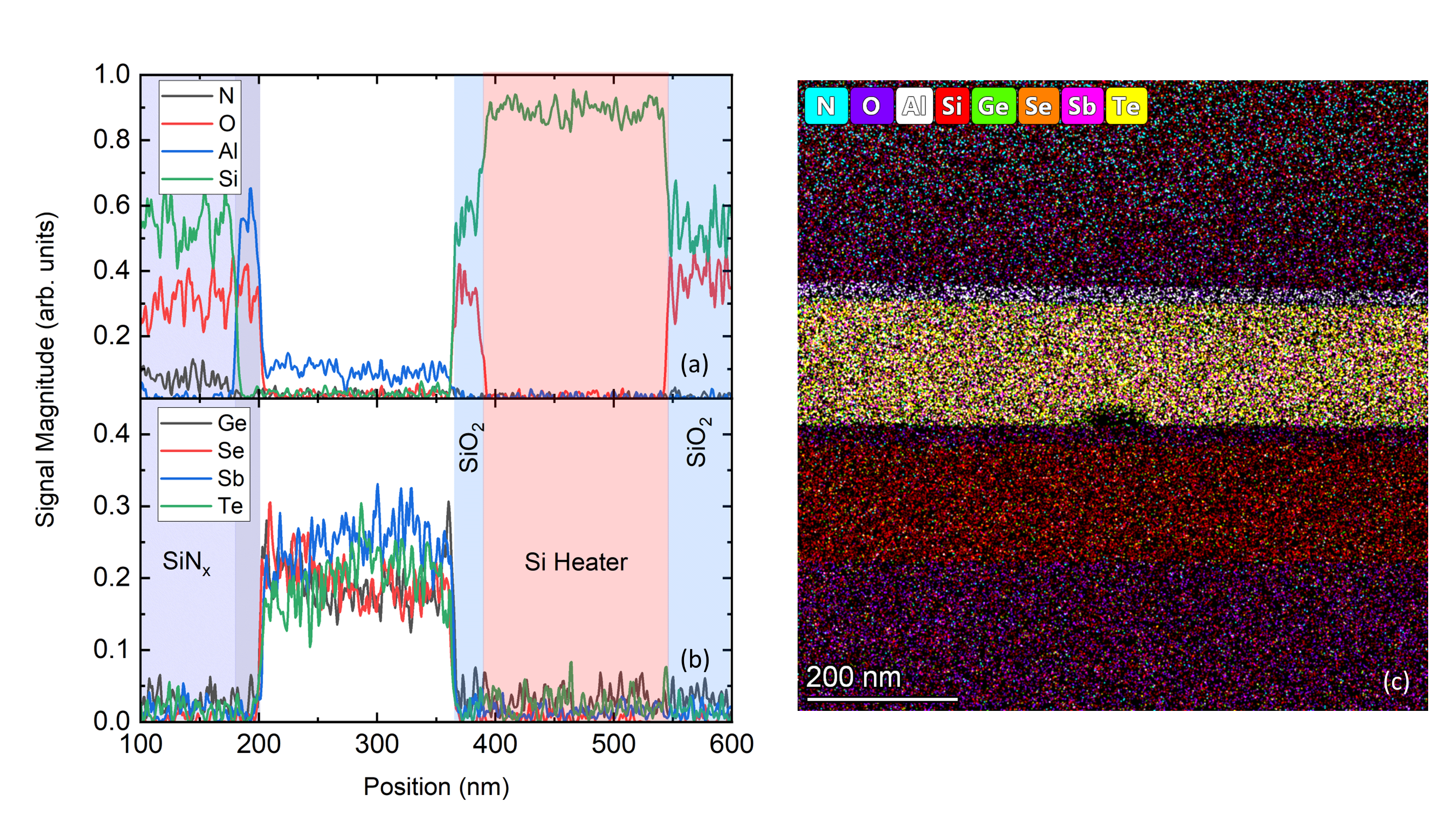}}
\caption{Layers composition in the device after 375 cycles.
\label{fig_EDS_02}}
\end{figure*}

From EDS mapping of the thin film PCM device, it can be seen that Ge is deposited preferentially before Sb and Te during the thermal evaporation.Upon cycling, the reflectivity contrast increases during the burn-in cycles, most likely because this composition gradient along the thickness of the film is removed. EDS line scans highlight further that the as-fabricated film has a slightly heavier concentration of Se towards the bottom of the film vs the top (Fig. \ref{fig_EDS_01}), but the difference is not as stark as for Ge and Sb. 375 cycles is significantly beyond the range of the burn-in, with a relatively more flat composition profile being observed at this time point in the device (Fig. \ref{fig_EDS_02}).  


\clearpage

\section*{Supplementary Note 4}
\textbf{COMSOL simulation}

Joule heating in the PCM meta-atoms was performed in COMSOL\textregistered. The material properties of the PCM Ge$_2$Sb$_2$Se$_4$Te (GSST), Si, SiO$_2$, and SiN$_x$ can be found in Table \ref{tab:01}. The thermal conductivity for SiN$_x$ is assumed to be 2 $\frac{W}{mK}$ because the silicon nitride was deposited via reactive sputtering and it is reasonable to assume high defect density limiting heat transfer and the value would be in line with films deposited via similar methods \cite{Braun2021HydrogenH}. 
The density of the amorphous GSST phase is expected to be smaller than that of crystalline GSST, but in absence of measured results and for the purpose of approximation of the results via simulation, the density for GSST in amorphous and crystalline form are taken as equal.

The latent heat of fusion for GSST was assumed to be equal to that of Ge$_2$Sb$_2$Te$_5$ (GST) (147 $\frac{kJ}{kg}$ \cite{Aryana2023TowardDevicesb}) due to lack of data and similarity between the two compositions. To model the impact of the patterned structure on the heat transport, a smaller initial simulation with fully patterned structures was performed from which the effective anisotropic thermal conductivity and effective heat capacity of the film were extracted. Afterwards, the 200 $\mu$m $\times$ 200 $\mu$m device was simulated with a film with the effective values for conductivity and heat capacity. To see the impact of the metasurface in the device, actual patterned GSST and SiN$_x$ were placed at the points of interest (corner, middle of the edge and center of the heater), superseeding the effective thin film. This method was implemented due to the large computational requirements of running a fully patterned metasurface in a 3D electrically induced heating simulation with the appropriate spatial and temporal resolution. For boundary conditions, the silicon layer below the buried oxide was kept at constant room temperature. Convective cooling is negligible on the time scale of amorphization. 

\setlength{\tabcolsep}{.7em}
\begin{table}[htb]
\caption{\label{tab:01} Material properties for thermal simulations. Electric conductivities shown as N/A are assumed too low in comparison to the doped Si heater conductivity to impact the results. Unless otherwise specified, the material properties were taken from the COMSOL \textregistered V6.1 material database.}
  \centering
  \begin{tabular}{@{}ccccc@{}}
    \toprule
    \toprule
        Property & $\sigma (\frac{S}{m})$&$\kappa (\frac{W}{mK})$ & $c_p (\frac{J}{kg K})$ & $\rho (\frac{kg}{m^3})$  \\  
        \bottomrule
        a-GSST & N/A \cite{Zhang2019BroadbandPhotonics}  & 0.2   & 270 \cite{Aryana2021SuppressedGe2Sb2Se4Te} & 5530  \\
        c-GSST  & N/A \cite{Yuan2023ElectrodePulse, Zhang2019BroadbandPhotonics} &  0.5  & 325 \cite{Aryana2021SuppressedGe2Sb2Se4Te} & 5530 \cite{Aryana2021SuppressedGe2Sb2Se4Te} \\
        i-Si 	& N/A& 130      & 700 & 2329  \\ 
        n++ Si 	& 1.34-0.80$\times$10$^5$ & 130         & 700 & 2329  \\
        SiO$_2$	& N/A&  1.38            & 703 & 2203  \\
        SiN$_x$ & N/A&  2 \cite{Braun2021HydrogenH}& 700 & 3100  \\
    \bottomrule
    \bottomrule
  \end{tabular}
\end{table}

\begin{figure*}[h!]
\centering
{\includegraphics[width=  6.5 in]{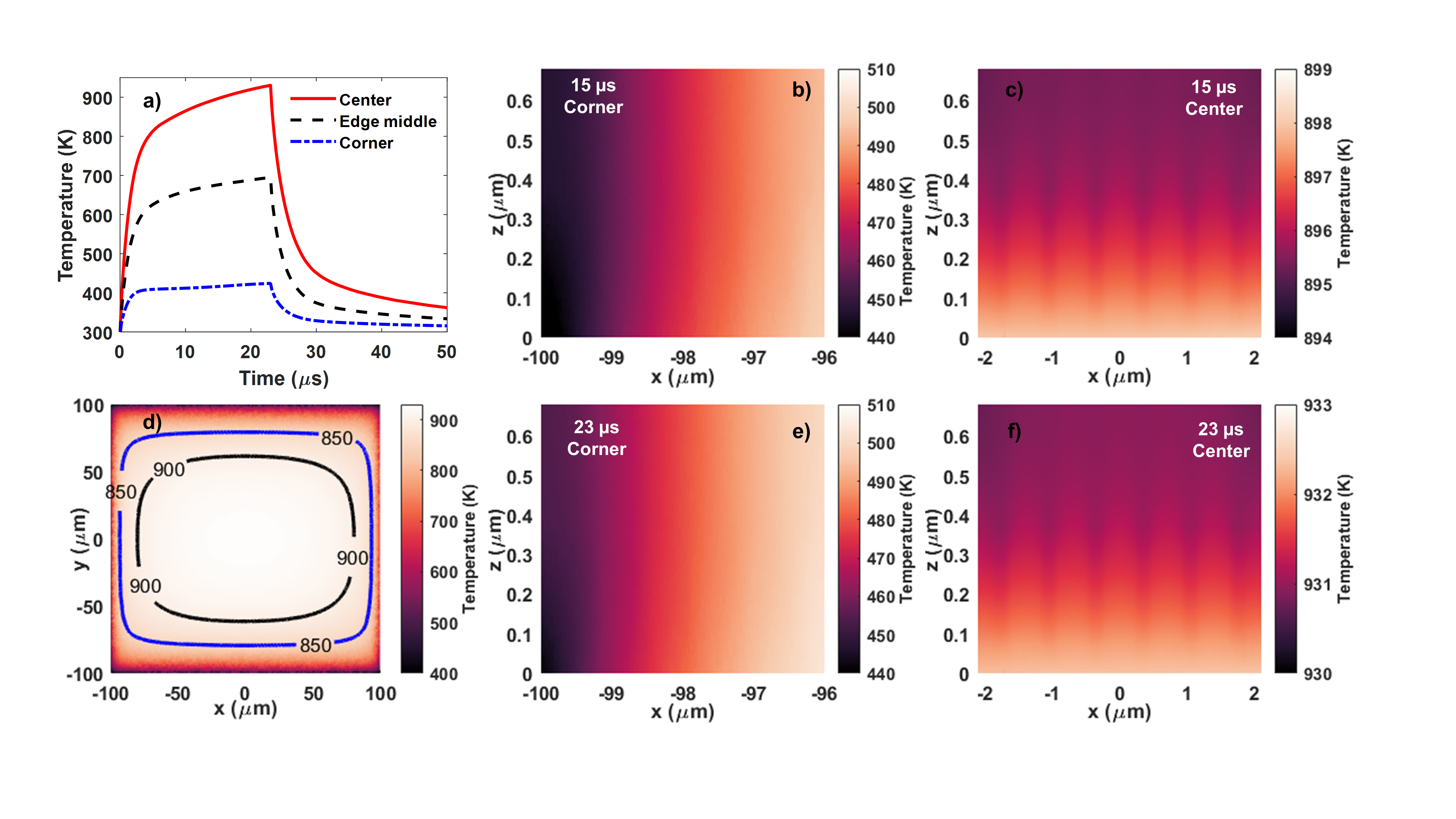}}
\caption{Transient thermal simulations of the amorphization of the heater for the meta-surface. (a) Temperature plots from the center, middle of the edge and corner of the device. (b, e) Cross sections through the meta-atoms placed at the corner and  (c,f) center of the heater after (b,c) 15 $\mu s$ and (e,f) 23 $\mu s$ throughout the application of the pulse, along with (d) a map showing the temperature in the plane of the device. 
\label{fig_COMSOL_main}}
\end{figure*}

During the amorphization pulse, the regions at the boundary of the heater heat to a significantly smaller extent as in comparison to the center of the heater due to heat extraction by the intrinsic silicon of the rest of the chip and, more importantly, a lower rate of local heat generation. The cooling rate across the three regions is primarily dictated by the buried oxide thickness which acts as a thermal resistor in series with the device and PCM structure (Fig. \ref{fig_COMSOL_main} a). When comparing the metasurface regions simulated at the corner vs the center of the heater, surprisingly, the patterning of the structure does not show significant impact in terms of the temperature distribution. As the pulse is applied, the temperature difference across the PCM plus nitride protective layer is less than 5 K after 15 $\mu$s and 3 K after 23 $\mu$s, meaning that out of plane thermal gradients are relatively low in comparison to in-plane thermal gradients (Fig. \ref{fig_COMSOL_main} c,f). There is a significant temperature gradient at the corner of the boundary of the heater. This gradient is due to the lower rate of heat generation, and the impact of the meta-atoms placed at the corner is virtuallyinconsequential. It is hypothesized that an important component of this is the low thickness of the stack and relative proximity in thermal conductivity of GSST and SiN$_x$ used in the simulations. Potentially, if the protective layer had a thermal conductivity several orders of magnitude above that of GSST and the analysis was performed before significant heat diffusion into the PCM had time to take place, the heterogeneity of the structure would show up in the results. A comparison of the temperature across the corner region between the two time points shows the heat front propagated due to the applied pulse (Fig. \ref{fig_COMSOL_main} b,d). When analyzing the temperature in the plane of the heater, a central relatively uniform region of around 150 $\mu$m $\times$ 100 $\mu$m can be seen, with a maximum drop of around 30 K center to edge. This is in contrast with the edges of the heater which after 23 $\mu$s are close to 400 K, showing again the relatively large thermal gradient forming close to the heater boundaries. 

\newpage

\section*{Supplementary Note 5}

\begin{figure*} [h!]
\centering
    {\includegraphics[width=  6.5 in]{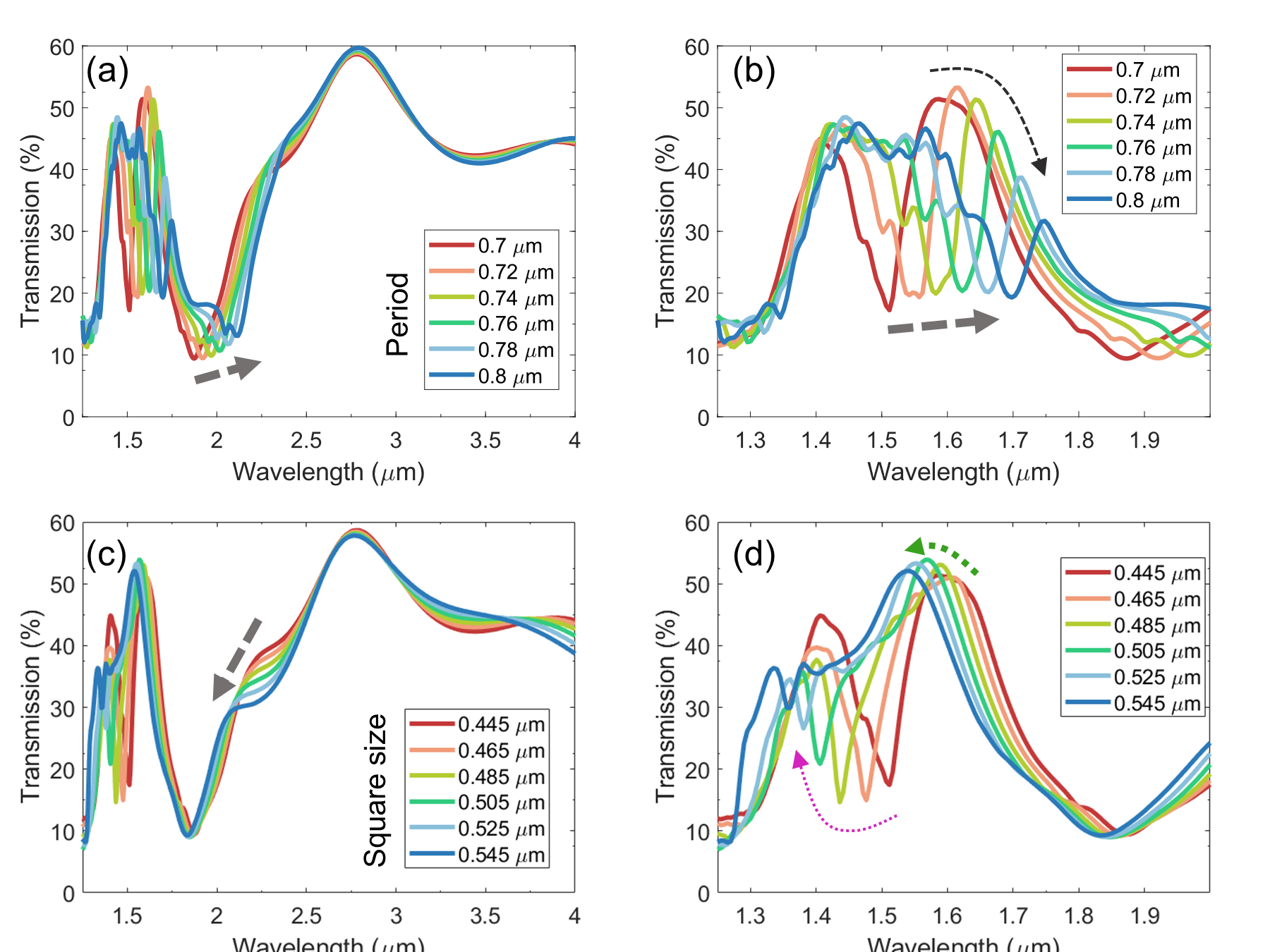}}
\caption{ RCWA Simulations of fully amorphous devices with varying periods (a-b) (445 nm hole) and varying square hole sizes (c-d) (700 nm period) highlighting the red shift due to effective increase in refractive index of the structure for larger periods, or the blue-shift for larger hole sizes leading to lower effective index. (b) and (d) are zoom-ins of the (a) and (c) respectively for better visualization of the peak shifts. The arrows are guides to the eye, following the variation of the targeted parameters.   
\label{fig_RCWA_Examples} }
\end{figure*}

Several RCWA simulations at 0\% crystalline fraction on the same type of structure as the fabricated / simulated one in the main manuscript are shown in Fig. \ref{fig_RCWA_Examples}. The fully amorphous state was chosen for better clarity of the impact and trends on the features in the spectrum, despite the likely inability to reach the full amorphous state in the fabricated devices.
With larger periods at constant hole size in the structure, we notice a red shift of the spectral features (i.e. transmission peaks and valleys). This can be associated to the effective increase in the fill factor of the structure with the high index material (i.e. the PCM), leading to this transformation\cite{guidedModeRes2002Fan}.
Conversely, when keeping the same period and increasing the square size, the region with PCM located in the meta-atom decreases, and thus the effective index of the layer, leading to an effective blue-shift of the observed resonances in the system (Fig. \ref{fig_RCWA_Examples}(d)). Fig. \ref{fig_RCWA_Examples}(b) shows a large set of local peaks, especially at 800 nm period. This may be caused due to the many interfaces in the structure leading to many regions that can create resonances in the meta-atom. It is possible that some of these features may be lost in fabricated devices in the case of significant roughness induced in the meta-atom during RIE or during the reactive sputtering, beyond the limitations in reaching full amorphization in the device. The region beyond 2.5 $\mu$m shows little impact from the variation of these parameters for the chosen simulated regime. A decrease in the amplitude of the resonance/dip in transmission can be observed in Fig. \ref{fig_RCWA_Examples}(d) as it goes from 1.5 $\mu$m to 1.35 $\mu$m. This may be associated with an increase in the extinction coefficient of GSST, leading to lower quality factors and a decrease in the impact of the meta-atom on the transmission modulation.

\clearpage
\bibliography{references.bib}